\catcode`\@=11

\def\citen#1{\if@filesw \immediate\write \@auxout {\string\citation{#1}}\fi%
\@tempcntb\m@ne \let\@h@ld\relax \def\@citea{}%
\@for \@citeb:=#1\do {\@ifundefined {b@\@citeb}%
    {\@h@ld\@citea\@tempcntb\m@ne{\bf ?}%
    \@warning {Citation `\@citeb ' on page \thepage \space undefined}}%
    {\@tempcnta\@tempcntb \advance\@tempcnta\@ne
    \setbox\z@\hbox\bgroup\ifcat0\csname b@\@citeb \endcsname \relax
    \egroup \@tempcntb\number\csname b@\@citeb \endcsname \relax
    \else \egroup \@tempcntb\m@ne \fi \ifnum\@tempcnta=\@tempcntb
    \ifx\@h@ld\relax \edef \@h@ld{\@citea\csname b@\@citeb\endcsname}%
    \else \edef\@h@ld{\hbox{--}\penalty\@highpenalty
    \csname b@\@citeb\endcsname}\fi
    \else \@h@ld\@citea\csname b@\@citeb \endcsname \let\@h@ld\relax \fi}%
\def\@citea{,\penalty\@highpenalty\hskip.13em plus.13em minus.13em}}\@h@ld}
\def\@citex[#1]#2{\@cite{\citen{#2}}{#1}}%
\def\@cite#1#2{\leavevmode\unskip\ifnum\lastpenalty=\z@\penalty\@highpenalty\fi%
  \ [{\multiply\@highpenalty 3 #1%
  \if@tempswa,\penalty\@highpenalty\ #2\fi}]}   %
\makeatother 

\def\abar          {{\liefont h}}

\def\aff           {affine Lie algebra}
\def\afp           {^\natural}
\def\alg           {algebra}
\def\alphab        {{\bar\alpha}}

\def\alrkg         {\llb\Frac{(\alphab,\Lambdab+\rhob)}{\kV+\gV}\lrb}
\def\asac          {as a consequence}
\def\auto          {automorphism}
\def\B             {\mbox{$B$}}
\def\barB          {{\overline{\cal H}^{\sss(\mn-2)}}}
\def\BarB          {{\overline{\cal H}^{\sss(\mn-1)}}}
\def\barchi        {\raisebox{.15em}{$\bar\chi$}}
\def\barG          {G}
\def\barH          {T}

\def\barW          {{\overline W}}
\def\be            {\begin{equation}}
\def\bearl         {\begin{array}{l}}
\def\bearll        {\begin{array}{ll}}
\def\bearlll       {\begin{array}{lll}}
\def\betab         {{\bar\beta}}
\def\bfe           {{\bf1}}
\def\bL            {L}  
\def\bla           {block algebra}

\def\bwb           {Borel\hy Weil\hy Bott }

\def\calb          {{\cal B}}
\def\calbb         {\bar{\cal B}}

\def\calh          {{\cal H}}
\def\calhb         {\overline{\cal H}}

\def\calo          {{\cal O}}

\def\cb            {chiral block}
\def\cB            {\mbox{\sf B}}

\def\cft           {conformal field theory}
\def\Cft           {Conformal field theory}
\def\cfts          {conformal field theories}

\def\chib          {{\bar\chii}}
\def\chibarx       {\chibarX\,}
\def\chibarX       {\chibl2\chibl3{\cdots}\,\chibl{\mn-1}}
\newcommand\chibl[1]{\bar\chii_{\Lambdab_{#1}}}
\def\chii          {{\raisebox{.15em}{$\chi$}}}
\newcommand\chil[1]{\chii_{\Lambda_{#1}}}
\def\chila         {\chii_\Lambda^{}}
\def\chilb         {\bar\chii_\Lambdab^{}}

\def\Circ          {\,{\circ}\,}
\newcommand\coi[2] {\lfloor #1\rfloor^{}_{#2}}
\newcommand\coI[2] {\lfloor #1\rfloor_{#2}}
\newcommand\coidual[2]{\lfloor #1\rfloor\dual_{#2}}
\def\coin          {co-in\-va\-ri\-ant}
\def\Coin          {Co-in\-va\-ri\-ant}
\def\complex       {{\dl C}}
\def\Complex       {$\dl C$}

\def\congto        {\stackrel\cong\to}
\def\cS            {Chern\hy Si\-mons\ }
\def\csa           {Cartan subalgebra}

\def\df            {\,{:=}\,}

\def\dg            {\rmd\mu_G}

\def\dim           {{\rm dim}\,}
\def\dl            {\mathbb }

\def\dsum          {\displaystyle\sum}
\def\dstyle        {\displaystyle}
\def\dt            {\rmd t} 
\def\dual          {^\star}

\def\ee            {\end{equation}}
\def\eE            {{\rm e}}
\def\eear          {\end{array}}

\def\eps           {\epsilon}

\def\epsw          {\epsilon(w)}
\def\epswb         {\epsilon(\wbar)}

\def\eq            {\,{=}\,}
\newcommand\erf[1] {(\ref{#1})}
\newcommand\Erf[2] {(\ref{#1#2})}
\def\eul           {\mbox{\sc x}}
\def\Exp           {\mbox{\sc Exp}}

\def\ezm           {E^\ttA\otim t^{-1}}
\def\ezp           {E^{-\ttA}\otim t}
\newcommand\fc[4]  {{#1}\hspace{-#2em}\raisebox{#3em}{$\scriptstyle\circ$}
                   \hspace{#4em}}

\newcommand\fhto[1]{\stackrel{f^{#1}_\h}{\longrightarrow}}
\def\findim        {finite-dimensional}

\def\fpi           {f_{p_i}}  

\def\fqj           {f_{q_j}}  
\newcommand\Frac[2]{\mbox{\large$\frac{#1}{#2}$}}

\newcommand\fto[1] {\stackrel{f^{#1}_{}}{\longrightarrow}}

\def\futnote#1     {\footnote{~#1}\ }

\def\g             {\mbox{$\liefont g$}}

\def\gb            {\mbox{$\bar\gM$}}
\def\gbar          {\mbox{$\bar\gM$}}
\def\gB            {{\bar\gM}}
\def\gbo           {\mbox{$\gB^{}_\circ$}}
\def\gbodual       {\mbox{$\gB^\star_\circ$}}

\def\gcP           {\mbox{$\gB(\Pem)$}}
\def\gcPM          {{\gB(\Pem)}}
\def\gh            {t}

\def\gl            {\mbox{$\bar\gM_{\rm loop}$}}

\def\gm            {\mbox{$\gM^{}_-$}}
\def\gM            {{\liefont g}}

\def\gminus        {{\gM_{}^-}}

\def\gmz           {\gM_{\vec\zeta}\!\!\raisebox{.5em}{$\sss(m)$}}
\def\gmzu          {\gM_{\vzvu}\hsp{-1.1}\raisebox{.5em}{$\sss(m)$\hsp{.3}}}
\def\go            {\mbox{$\gM^{}_\circ$}}
\def\godual        {\mbox{$\gM^\star_\circ$}}

\def\gp            {\mbox{$\gM^{}_+$}}

\def\gplus         {{\gM_{}^+}}

\def\gt            {{\liefont z}}
\def\gT            {$\liefont z$}
\def\gtc           {{\hat\gt}}

\def\gUU           {{\gB(U)}}
\def\gv            {{\rm h}}
\def\gV            {{\rm h}}
\def\GV            {{\rm h}}
\def\h             {{\liefont h}}
\def\H             {${\liefont h}$}

\def\hb            {\bar h}
\def\hh            {\mbox{$\HL1\!{\otimes}\barB$}}
\def\hhc           {\mbox{$(\HL1\!{\otimes}\barB)^{\overline{\phantom w}}$}}
\def\hhd           {\mbox{$(\HL1\!{\otimes}\barB)_{\phantom|}\dual$}}
\def\hhD           {\mbox{$\LlB\HL1\!{\otimes}\barB\LrB_{}\dual$}}
\def\hhdpsi        {\mbox{$(\HL1\!{\otimes}\barB)\dual_{\!(\hat\psi)}$}}
\def\hhe           {{\kV+\gV}}
\def\hhE           {{(\kV+\gV)}}

\def\hil           {\mbox{$\cal H$}}
\def\hiL           {{\cal H}}
\def\hilb          {\mbox{$\bar{\cal H}$}}
\def\hilt          {\tilde{\cal H}}
\def\hl            {\mbox{${\cal H}_\Lambda$}}

\newcommand\HL[1]  {{{\cal H}_{\Lambda_{#1}}^{}}}
\def\hlb           {\mbox{$\bar{\cal H}_{\Lambdab}$}}
\def\hli           {\mbox{${\cal H}^{}_{\Lambda_i}$}}
\def\hlib          {\mbox{$\bar{\cal H}^{}_{\Lambdab_i}$}}
\def\hll           {\mbox{\large$\left|\frac{(\kV+\GV)\bL^{\sss\vee}}
                   {\bL^{\rm w}_{}}\right|$}\,}
\def\hllw          {\mbox{\large$\frac1{|\overline W|}\,\left|\frac{(\kV+\GV)
                   \bL^{\sss\vee}}{\bL_{\rm w}^{}}\right|$}\,}

\def\hlm           {{\tilde{\cal H}_{\Lambda_\mn}}}
\def\hlw           {highest and lowest weight}
\newcommand\HM[1]  {{{\cal H}_{\mu_{#1}}^{}}}
\def\hmb           {\mbox{$\bar{\cal H}_{\Lambdab'}$}}
\def\hsa           {horizontal subalgebra}
\newcommand\hsp[1] {\mbox{\hspace{#1 em}}}
\def\hvl           {\mbox{${\cal H}_{\vec\Lambda}\!\!\!\raisebox{.5em}
                   {$\sss(m)$}$}}
\def\HVL           {${\cal H}_{\vec\Lambda}^{\sss(m)}$}
\def\hvz           {\mbox{${\cal H}_{\vec\Lambda}\!\!\!\raisebox{.5em}
                   {$\sss(2)$}$}}
\def\hw            {highest weight}
\def\hwm           {highest weight module}

\def\hwv           {highest weight vector}
\def\hy            {$\mbox{-\hspace{-.66 mm}-}$}
\def\id            {\mbox{\sl id}}
\def\ii            {{\rm i}}
\def\ihwm          {irreducible highest weight module}
\def\Im            {\mbox{\em Im}\,}
\def\infdim        {infinite-dimensional}
\def\iN            {\,{\in}\,}
\def\inonetom      {\iN\{1,2,...\,,\mn\}}
\def\intG          {\int_G\!\dg\,}
\def\inthwm        {integrable highest weight module}
\def\intT          {\int_T\!\dt\;}

\def\irmod         {irreducible module}

\def\j             {\imath}
\def\jm            {{j_{\rm max}}}
\def\jM            {{(j-1)}}
\def\jo            {{(0)}}
\def\jj            {{(j)}}
\def\Ker           {\mbox{\em Ker}\,}
\def\kerl          {{\cal K}_{\Lambda}}

\def\klo           {{\kV\Lambda_{(0)}}}

\def\kma           {Kac\hy Moo\-dy algebra}
\def\kpf           {Kac\hy Peterson formula}
\def\Kt            {\hat K}
\def\kv            {{\rm k}}
\def\kV            {{\rm k}}

\long\def\labl#1   {\label{#1}\ee \ifnum\draftcontrol=1
                   \mbox{ }\\[-12 mm]\query{#1}\\[5 mm] \fi}
\long\def\Labl#1#2 {\label{#1#2}\ee\ifnum\draftcontrol=1
                   \mbox{ }\\[-12 mm]\query{#1#2}\\[5 mm] \fi}
\def\lambdab       {{\bar\lambda}}
\def\Lambdab       {{\bar\Lambda}}
\def\Lambdam       {{\Lambda_m}}

\def\Ldots         {,...\,,}
\def\lhs           {left hand side}

\def\lie           {Lie algebra}
\def\Lie           {Lie group}
\def\liefont       {\mathfrak }
\def\liematrixfont {\mathfrak }
\def\lLb           {\mbox{\large[}}
\def\llb           {\mbox{\large(}}

\def\Llb           {\mbox{\Large(}}
\def\LlB           {\mbox{\LARGE(}}
\def\lm            {l}
\def\lo            {\Lambda_{(0)}}

\def\lRb           {\mbox{\large]}}
\def\lrb           {\mbox{\large)}}

\def\Lrb           {\mbox{\Large)}}
\def\LrB           {\mbox{\LARGE)}}

\def\Lrkg          {(\Frac{\Lambdab+\rhob}{\kV+\gV})}
\def\LrkG          {\!(\Frac{\Lambdab+\rhob}{\kV+\gV})}

\def\lv            {\mbox{$\bL^{\!\Vee_{}}_{\phantom|}$}}
\def\lV            {\bL^{\!\vee_{}}_{{}^{}}}
\def\lVV           {\bL^{\!\Vee_{}}_{\phantom|}{}\!}
\def\lw            {\mbox{$\bL_{\rm w}$}}
\def\lW            {\bL_{\rm w}}

\def\mapstO        {\,{\mapsto}\,}
\def\mi            {\,{-}\,}
\def\Mid           {\,{\mid}\,}
\def\mm            {{m-1}}
\def\mn            {{m}}

\def\mub           {{\bar\mu}}
\newcommand\mult[2]{[ #1 \,{:}\, #2]}

\def\nE            {\,{\not=}\,}
\def\nealg         {{\tilde\gM^{-}}}

\def\neh           {{\tilde {\cal H}}}
\def\nehe          {{\tilde {\cal H}_{\sss(1)}}}
\def\nehz          {{\tilde {\cal H}_{\sss(2)}}}
\newcommand\NN[1]  {N_{\!\Lambda_1;\Lambdab_2...\Lambdab_{\mn-1};#1}}
\def\NNL           {\NN{\Lambda_\mn^+}}
\def\nnu           {\nu}

\def\onedim        {one-dimen\-sional}

\def\onetom        {1,2,...\,,\mn}
\newcommand\oo[3]  {{{#1}^{\mskip-#3 mu\raise #2 pt\hbox{$\scriptstyle\circ$}}}}
\def\oog           {\oo{\gM}{1.6}{10.3}}
\def\op            {^+_{\phantom|}}

\def\opluS         {\,{\oplus}\,}
\def\ot            {\raisebox{.07em}{$\scriptstyle\otimes$}}
\def\oT            {\,\ot\,}
\def\otim          {\ot}
\def\otimesc       {\otimes_\complex^{}}
\def\otimeS        {\,{\otimes}\,}
\def\otimES        {{\otimes}\,}

\def\pbw           {Poin\-ca\-r\'e\hy Birk\-hoff\hy $\!$Witt theorem}
\def\pe            {\mbox{${\dl P}^1$}}
\def\pE            {{{\dl P}^1_{}}}
\def\Pem           {{{\dl P}^1_{\!\!\sss(\mn)}}}

\def\pii           {\pi{\rm i}}
\def\pk            {{{\rm P}_{\!\kV}}}
\def\pl            {\,{+}\,}
\def\PL            {\mbox{${\rm P}_{\!\Lambda}$}}
\def\PLP           {\mbox{${\rm P}_{\!\Lambda'}$}}
\def\pM            {{{\liefont p}_{}^+}}

\def\rank          {{\rm rank}\,}
\def\reals         {{\dl R}}
\def\Reals         {$\dl R$}
\def\rep           {rep\-re\-sen\-ta\-ti\-on}
\def\Rep           {Representation}
\def\Res           {{\rm Res}}
\def\resi          {{\rm Res}_{p_i}}
\def\resinf        {{\rm Res}_{\infty}}
\def\resp          {respectively}
\def\rest          {{\cal J}}

\def\rhob          {\bar\rho}
\def\rhs           {right hand side}
\def\rvl           {\mbox{$R_{\vec\Lambda}\!\!\!\raisebox{.5em}
                   {$\sss(m)$}$}}
\def\rvlu          {\mbox{$R_{\vec\Lambda;\vzvu}\hsp{-1.85}\raisebox{.55em}
                   {$\sss(m)$}$\hsp{.85}}}
\def\rvlz          {\mbox{$R_{\vec\Lambda;\vec\zeta}\hsp{-1.05}\raisebox{.55em}
                   {$\sss(m)$}$\hsp{.1}}}
\def\Rvlz          {{R^{\sss(m)}_{\vec\Lambda;\vec\zeta}}}
\def\rmd           {{\rm d}}

\def\scbs          {spaces of chiral blocks}
\def\scs           {\scriptstyle}

\newcommand\sref[1]{section \ref{s.#1}}
\newcommand\sect[1]{\section{#1}\setcounter{equation}{0}}
\newcommand\Sect[2]{\sect{#1}\label{s.#2} \ifnum\draftcontrol=1 \query{s.#2}\fi}
\def\Setminus      {\,{\setminus}\,}

\def\sln           {\mbox{$\liematrixfont{sl}(N)$}}

\def\sltwo         {\mbox{$\liematrixfont{sl}(2)$}}
\def\slz           {\mbox{SL$(2{,}\zet)$}}

\def\smat          {$S$-matrix}
\def\soc           {\mbox{{\sl soc}$_\pm$}}
\def\socm          {\mbox{{\sl soc}$_-$}}

\newcommand\ssref[1]{subsection \ref{ss.#1}}
\def\sss           {\scriptscriptstyle}
\def\subseT        {\,{\subset}\,}

\newcommand\sumk[1]{\dsum_{#1\in\pk}}
\def\summi         {\dsum_{i=1}^m}
\def\summI         {\sum_{i=1}^m}

\newcommand\sumw[1]{\dsum_{#1\in W}}
\def\sumwlj        {\!\!\bigoplus_{\ \ \scs\wP\in\WP^{}_{\phantom{X}}\atop
                   \scs\ell(\wP)=j}\!\!\!}
\def\sumww         {\sum_{w\in W}}
\def\sumwwb        {\sum_{\wbar\in\Wb}}
\def\sumwwk        {\sum_{\wP\in\WP}}

\def\tilf          {{\tilde f}}
\def\tildeg        {{\tilde\gM}}
\def\Times         {\,{\times}\,}
\def\tlk           {{-\tau L_0{+}\zea K}}

\def\ttA           {{\bar\theta}}
\def\ttV           {{\bar\theta^\Vee}}
\newcommand\twobrac[1]{(\!(#1)\!)}
\def\twodim        {two-di\-men\-si\-o\-nal}
\def\U             {{\sf U}}
\def\uaff          {untwisted affine Lie algebra}

\newcommand\uep[1] {u_{\sss(1),#1}^+}
\newcommand\ueP[1] {\tilde u_{\sss(1),#1}^+}

\def\ugmi          {\mbox{$\U(\gminus)$}}
\def\Up            {\U^+}

\def\uvc           {{\cal X}}
\def\uvcb          {\bar\uvc}

\newcommand\uzp[1] {u_{\sss(2),#1}^+}
\newcommand\uzP[1] {\tilde u_{\sss(2),#1}^+}

\def\Vee           {{\scriptscriptstyle\vee}}
\def\Ver           {{\cal V}}

\def\Verl          {{\cal P}_{\!\Lambda}}
\def\Verm          {{\cal P}}
\def\vermod        {Verma module}
\newcommand\version[1] {\ifnum\draftcontrol=1 \typeout{}\typeout{#1}\typeout{}
                   \vskip3mm \centerline{\fbox{{\tt DRAFT -- #1 -- }
                   {\small\draftdate}}}
                   \vskip3mm \fi}
\def\vGu           {\Gamma_{\!\vec u}}

\def\vl            {\mbox{$v^{}_\Lambda$}}

\def\VV            {{\cal P}}
\def\vzvu          {\vec\zeta\circ\vec u}
\def\Wb            {\overline W}
\newcommand\wb[1]  {w(\bar{#1}+\bar\rho)-\bar\rho}

\def\wbar          {{\bar w}}
\def\wcf           {Weyl character formula}
\def\what          {{\hat w}}

\def\wkcf          {Weyl\hy Kac character formula}
\def\wkr           {{\wp}}
\def\wkR           {{\wP}}
\def\Wkr           {{\Wp}}

\def\wmb           {\overline{w\mbox{{\footnotesize(}$\mu${\footnotesize)}}}}

\def\wmax          {\bar w_{{\rm max}}}

\def\wp            {\fc w{.5}{.48}{.1}}
\def\wP            {\fc w{.37}{.32}{.1}}
\def\Wp            {\fc W{.7}{.78}{.2}}
\def\WP            {\fc W{.55}{.52}{.2}}
\def\wrt           {with respect to }

\def\wrtt          {with respect to the }

\def\WZW           {Wess\hy Zu\-mi\-no\hy Wit\-ten}
\def\wzwm          {WZW model}
\def\wzwt          {WZW theory}
\def\wzwts         {WZW theories}
\def\xb            {{\bar x}}

\def\xmn           {\xb\ot t^{-n}}
\def\xn            {\xb\ot t^n}

\def\yb            {{\bar y}}

\def\z             {\zeta}
\def\zea           {\varpi}
\def\zet           {{\dl Z}}

\def\zetminus      {\mbox{$\zet_{<0}$}}

\def\zetplus       {\mbox{$\zet_{>0}$}}

\def\zh            {\zpi h}
\def\zhb           {\zpi\hb}
\def\zpi           {2\pi{\rm i}}

\newcommand\zz[1]  {(z\mi z_{#1})}

\def\draftdate{\number\month/\number\day/\number\year\ \ \ \hourmin }

\global\def\draftcontrol{0}
\catcode`\@=12

\documentclass[12pt]{article} \usepackage{amssymb,amsfonts}

\begin{document}

\begin{flushright}  {~} \\[-15 mm]  {\sf hep-th/9707069} \\[1mm]
{\sf May 1998 (revised)} \\[1 mm] {\sf (original version: CERN-TH/97-152)}
\end{flushright} 

\begin{center} \vskip 13mm
{\Large\bf A REPRESENTATION THEORETIC APPROACH}\\[2.8mm]
{\Large\bf TO THE WZW VERLINDE FORMULA }\\[16mm]
{\large J\"urgen Fuchs} \\[3mm]
DESY\\[.6mm] Notkestra\ss e 85, \ D -- 22603~~Hamburg\\[11mm]
{\large Christoph Schweigert} \\[3mm] CERN \\[.6mm] CH -- 1211~~Gen\`eve 23
\end{center}
\vskip 20mm

\begin{quote}{\bf Abstract}\\
By exploring the description of chiral blocks in terms of co-invariants,
a derivation of the Verlinde formula for WZW models is obtained which 
is entirely based on the representation theory of affine Lie algebras. 
In contrast to existing proofs of the Verlinde formula, this approach 
works universally for all untwisted affine Lie algebras. As a by-product 
we obtain a homological interpretation of the Verlinde multiplicities
as Euler characteristics of complexes built from invariant tensors of 
finite-dimensional simple Lie algebras.\\
Our results can also be used to compute certain traces of automorphisms 
on the spaces of chiral blocks.
\end{quote}
\newpage


\Sect{Introduction and summary}I

Spaces of chiral blocks are \findim\ vector spaces that arise naturally 
in the study of moduli spaces of flat connections over complex curves;
therefore they emerge in various contexts in physics and mathematics.
These vector spaces form a vector bundle over the moduli space of
smooth projective complex curves with marked points.
The aim of this paper is to further elucidate the 
structure of such spaces by using the \rep\ theory of affine \lie s. 

In physics, spaces of \cb s appear in the following guises.
In three-dimensional topological Chern\hy Simons gauge theories 
with space-time equal to the product of \Reals\ (describing time) and a
complex curve, they arise as the spaces of physical states
that are obtained when quantizing the theory in the temporal gauge. 
In \twodim\ \cft\ the chiral blocks are the basic constituents
of correlation functions, which are the quantities of prime interest in 
any quantum field theory.
More precisely, correlation functions on closed orientable Riemann surfaces are 
obtained as bilinear combinations of \cb s, while correlators on surfaces
that have boundaries and\,/\,or are unorientable can be expressed in terms
of linear combinations of \cb s. Correlators satisfying boundary conditions 
that correspond to the presence of so-called $D$-branes are expressible in 
terms of \cb s, or close relatives thereof, as well.

In this paper we consider the spaces of \cb s which are associated to
WZW (\WZW) \cfts. A WZW theory is specified by
the choice of a \findim\ simple \lie\ \gb\ and a positive integer \kv. In this 
case of our interest each of the marked points of the curve is labelled with 
an integrable weight at level $\kv$ of the untwisted affine \lie\ \g\ that is
associated to \gb\ via the loop construction. The \cb s of \wzwts\ 
are also of interest in algebraic geometry. 
Namely, the space of WZW \cb s can be interpreted as the space of holomorphic 
sections in the \kv-th tensor power of a line bundle over the
moduli space of flat \gb-connections over the curve. Chiral WZW blocks can
therefore be regarded as non-abelian generalizations of theta functions. In this
paper we use the fact that the space of \cb s can be described in terms of 
\coin s of certain integrable modules over \g\ (this characterization of 
\cb s will be reviewed in \ssref d). Based on this description we can 
apply tools from the \rep\ theory of the affine \lie\ \g\ to study the 
structure of \cb s.  As a crucial ingredient we will introduce a suitable 
central extension of the so-called \bla\ $\gb({\cal C})$, which by definition 
(for details see subsection \ref{gf}) consists of the \gb-valued algebraic 
functions on the (punctured) curve $\cal C$.

The most fundamental information about a sheaf of \cb s is its rank; there 
exist closed expressions for this quantity, which are commonly referred to as
{\em Verlinde formul\ae\/} \cite{verl2,sorg}. Owing to factorization theorems 
\cite{tsuy,falt,beau,tele2}, the problem of computing these numbers can be
reduced to the case of a curve of genus zero. Accordingly it will be assumed 
throughout this paper that the genus is zero. 
The main result of this paper is a purely \rep-theoretic argument
for deriving the Verlinde formula for chiral blocks at genus zero.
In contrast to existing algebraic proofs of the Verlinde formula (which we will 
briefly list in \sref S) we work as long as possible in the framework 
of \infdim\ \lie s. This enables us to obtain the Verlinde formula in a 
uniform manner for {\em all\/} choices of the underlying \findim\ simple \lie\ 
\gb, including the cases of $\gb=F_4,\,E_6,\,E_7,\,E_8$ for which no rigorous
algebraic proof had been known so far. Another advantage of our approach
is that we can derive the formula for an arbitrary number of marked points
without invoking (genus-preserving) factorization rules; this can be 
interpreted as an independent check of these factorization rules at fixed 
genus zero. As a further by-product we obtain a description of the spaces of 
\cb s in terms of a complex of invariant tensors of \gb\ with vanishing
Euler characteristic.

The structure of our approach is as follows. We first determine
the spaces of (genus zero) two-point blocks by making use of the explicit form
of the two-point \bla; this is done in \sref T. In \sref A
the problem for an arbitrary number of marked points is reduced to the
case of two marked points and to the calculation of certain branching rules.
We then derive an integral formula for the latter and use it to find an
integral formula for the dimension of the space of \cb s. 
When doing so, we have to make an assumption about 
the existence of a suitable completion of the modules. Finally,
in \sref F, we employ a generalized Poisson resummation rule to cast the
integral formula into the usual form of the Verlinde formula, i.e.\ as a finite 
sum over elements of the matrix $S$ that describes the modular transformation 
properties of the characters (which are the one-point blocks on the torus).

We conclude the paper in \sref S with some remarks which set our approach into 
the context of related work. In particular we present a complex of \coin s 
that characterizes the \cb s, which has the property that the vanishing of 
its Euler characteristic is equivalent to the Verlinde formula.
We also comment briefly on a possible extension to non-unitary theories.
In \sref C the necessary \rep-theoretic background is summarized.
Some technical aspects have been relegated to appendices.

\Sect{Chiral blocks as \coin s}C

\subsection{The \alg s \g\ and \gcP}\label{gf}

A basic ingredient in the definition of \cb s is
an \uaff\ \g. (In \cft, the semidirect sum of \g\ with the Virasoro \alg\
plays the r\^ole of the chiral symmetry \alg\ of WZW theories.) For the
purposes of this paper we regard \g\ as the centrally extended loop algebra
  \be  \gM:= \gB\otimes\complex\twobrac t \oplus \complex K  \,,  \labl g
where $\complex\twobrac t$ denotes the ring of Laurent series in some
indeterminate $t$ and \gb\ is a \findim\ simple \lie,
which is isomorphic to, and will be identified with, the horizontal (i.e.
zero mode) subalgebra of \g. The Lie bracket relations of \g\ read $[\xb\ot f,
\bar y\ot g]\eq[\xb,\yb]\ot fg + \kappa(\xb,\yb)\,\Res_0({\rm d}f\,g)\,K$ 
for $\xb,\yb\iN\gb$ and $f,g\iN\complex\twobrac t$ (here $\kappa$ denotes
the Killing form of \gb) and $[K,\xb\ot f]\eq0$,
i.e.\ $K\iN\gM$ is a central element. We also introduce the sub\alg s
  \be  \gplus:=\gB\otimes t\,\complex[[t]] \,, \qquad
  \gminus:=\gB\otimes t^{-1}\complex[t^{-1}]    \labl{gpm}
of \g\ ($\complex[t]$ and $\complex[[t]]$ denote polynomials and 
arbitrary power series in $t$, \resp);
then as a vector space \g\ can be decomposed as
  \be  \gM \,=\, \gminus \oplus\gB \oplus \complex K \oplus \gplus \,. \labl{gM}
The sub\alg s $\gM^\pm$ must not be confused with the maximal nilpotent 
subalgebras $\gM_\pm$ that appear in the triangular decomposition
\be  \gM \,=\, \gM_-\oplus\go\oplus\gM_+  \Labl1t
of \g\ into the \csa\ \go\ and the
nilpotent sub\alg s $\gM_\pm$ that correspond to the positive and negative
\g-roots, \resp; one has $\gM_\pm\cong \gM^\pm\opluS\gB_\pm$.

Concerning our characterization \erf g of \g, two remarks are in order.
First, apart from defining the grading that corresponds to the power of $t$,
in the present context the outer derivation $D=-L_0$ of the affine 
algebra will not play any particular r\^ole.
Accordingly we did not include $D$ in the definition \erf g, even though it
is e.g.\ needed in order for \g\ to possess a non-degenerate invariant
bilinear form and roots of finite multiplicity \cite{KAc3}.
Second, strictly speaking it is the sub\alg\ $\oog=\gB\otimes
\complex[t,t^{-1}] \oplus \complex K$ of \g\ that is generated
when one allows only for Laurent polynomials rather than arbitrary Laurent 
series which should be referred to as the affine \lie. However, 
as will be explained in the following subsection, we will only deal with
\g-\rep s for which 
every vector of the associated module (\rep\ space) is annihilated by
all but finitely many generators of the sub\alg\ $\oog{}_{}^+$ of $\oog$.
As a consequence, any such \rep\ of $\oog$ can be naturally promoted
to a \rep\ of the larger \alg\ \g. For the present purposes 
the distinction between \g\ and $\oog$ is therefore immaterial.
Note that unlike in the case of $\oog$, the sub\alg s
$\gplus$ and $\gminus$ of \g\ are not isomorphic; 
in particular, $\gminus$ is a sub\alg\ of $\oog$, while $\gplus$ is not.

The physical states of a WZW \cft\ can be completely described in terms of the
\rep\ theory of the affine \alg\ \g; e.g.\
the WZW primary fields $\Phi\,{\equiv}\,\Phi_\Lambda$ correspond to the \hwv s
of integrable \ihwm s \hl\ of \g. 
However, the description of chiral blocks also involves
another \infdim\ \lie, which we will call the {\em\bla\/} and
denote by \gcP. The \bla\ is the \lie\ of \gb-valued algebraic functions on the
(punctured) curve. More precisely, it is defined as follows.
To any open subset $U$ of the Riemann sphere
\pe\ one associates the ring ${\cal F}(U)$ of algebraic functions on 
$U$ and the vector space $\gUU:=\gB\otimeS{\cal F}(U)$; here and
below all tensor products are taken over the complex numbers, unless
stated otherwise. The open subset of interest to us
  \be  \Pem := \pe \setminus \{ p_1,p_2\Ldots p_\mn \}  \,,  \ee
where $\{p_i\Mid i\eq\onetom\}$ is a finite set of pairwise distinct
non-singular points on \pe. The points 
$p_i$ correspond to the positions of the WZW primary fields $\Phi_{\Lambda_i}$ 
whose correlation function is obtained from the chiral blocks we are interested 
in; they are called the {\em insertion points\/} or the {\em parabolic 
points\/}, and $\Pem$ is referred to as a punctured Riemann sphere.
The corresponding vector space
  \be  \gcP:= \gB\otimes{\cal F}(\Pem)  \Labl25
becomes a \lie\ when endowed with the natural bracket
  \be  [\xb\otim f,\bar y\otim g]:=[\xb,\yb]\oT fg
  \qquad{\rm for}\;\ \xb,\yb\iN\gB \;\ {\rm and}\;\ f,g\iN{\cal F}(\Pem) \,.
  \Labl xy
A basis $\calb$ of the \bla\ \gcP\ is given by
  \be  \calb = \calbb \times \llb \{z^0\} \cup \bigcup_{i=1}^m
  \{ \zz i^n\Mid n\iN\zetminus \} \lrb \,,  \ee
where $\calbb\eq\{\xb^a\Mid a\eq1,2\Ldots{\rm dim}\,\gB\}$ is a basis of
the \lie\ \gb, $z$ is the global coordinate of $\complex\subseT\pe$, and
$z_i$ are the values of $z$ at the insertion points $p_i$.

As a side remark we mention that the block algebra \erf{25} admits a natural
central extension, which we will employ in \sref T. A similar remark applies
to the block algebras that arise when \pe\ is replaced by some
Riemann surface of higher genus. The corresponding centrally extended
\lie s are known as higher genus affine \lie s or as generalized 
Krichever\hy Novikov algebras of affine type \cite{scHl1,shei3}.

\subsection{\g-modules}

For the applications we have in mind, we will need to consider modules
(\rep\ spaces) over the block algebra which come from modules over the affine 
\lie\ \g. Most of the \g-modules that we are interested in here 
share the specific properties that the action of
the \csa\ $\go$ can be diagonalized in such a way that the resulting weight
spaces are \findim\ when the full \csa\ (i.e.\ including the derivation $D$)
is considered, and such that each weight of the module can be obtained from
a finite set $\{\mu_\ell\}$ of weights by subtraction of a finite number of
positive \g-roots. The collection of all such \g-modules forms the objects of a
category, called the {\em category\/} $\calo$ (see e.g.\ Chap.\,9.1 of 
\cite{KAc3}); this category is closed under forming finite direct sums or 
tensor products, submodules and quotients. Every module $V$ in $\calo$ 
is in particular {\em restricted\/}, i.e.\ each element $v\iN V$ is annihilated
by the step operators for all but a finite number of positive \g-roots; 
moreover, the sub\alg\ $\gplus$ of \g\ acts locally nilpotently.

Among the modules in $\calo$ there are in particular the \hwm s $V$
for which by definition 
there exists a highest weight vector \vl\ which is annihilated by \gp, i.e.\
$\g_\alpha\vl\eq0$ for all positive \g-roots $\alpha$, which is an eigenvector
of \go, i.e.\ $h\vl\eq\Lambda(h)\vl$ for all $h\iN\go$, and for which the 
action of \gm\ yields the whole module, $\U(\gm)\vl= V$.
In particular, for \hwm s the set $\{\mu_\ell\}$ of distinguished weights 
contains only a single element, the \hw\ $\Lambda$. Every \g-module with \hw\ 
$\Lambda$ can be obtained as a suitable quotient from the {\em\vermod\/} 
$\Ver_\Lambda\df \U(\gM){\otimes_{\U(\gM^{}_+\oplus\gM^{}_\circ)}}\vl$,
which as a $\g_-$-module is isomorphic to the free module
$\U(\gM_-){\otimesc}\vl$.

The space of physical states of a (chiral) \wzwt\ is the direct sum of
integrable \ihwm s \hl\ of \g\ which all have one and the same eigenvalue of 
the central generator $K\iN\gM$, i.e.\ the same level. Throughout the paper we
therefore keep some fixed value \kv\ of the level. For integrability of 
the module \hl, the \hw\ $\Lambda$ must be dominant integral, which implies
that \kv\ must be a non-negative integer, and the horizontal projection
$\Lambdab$ of $\Lambda$ must lie in the set
  \be  \pk := \{ \Lambdab\iN\lw \Mid (\Lambdab,\bar\alpha^{(i)}_{})\,{\ge}\,0\;
  {\rm for\;all}\;i\eq1,2\Ldots{\rm rank}\,\gb,\;(\Lambdab,\ttA)\leq\kv \}
  \,, \Labl pk
i.e.\ belong to the integral weights in the closure
of the dominant Weyl alcove at level \kv\ (here \lw\ denotes the weight lattice
of the \hsa\ \gb, i.e.\ the lattice in \gbodual\ that is spanned
by the fundamental \gb-weights, and $\bar\alpha^{(i)}$ are the simple roots 
and $\ttA$ the highest root of \gb).
Note that $\pk$ is a finite set (e.g.\ at level 0 there is
only a single integrable module, the trivial \onedim\ module $\calh_0$
with \hw\ $\Lambda\eq0$).

In \cft\ terms, the WZW chiral blocks are the chiral constituents of the
$m$-point correlation functions
  \be  \cB_{\{\Lambda_i\},\{p_i\}} = \langle\,
  \Phi_{\Lambda_1}(p_1)\,\Phi_{\Lambda_2}(p_2)\cdots \Phi_{\Lambda_m}(p_m)
  \,\rangle  \ee
for primary WZW fields. These fields are associated with integrable \ihwm s 
of \g. As a consequence, when studying chiral blocks, one has to 
deal with a collection of \ihwm s \hli\ ($i\eq\onetom$) of \g\ which are at
level \kv\ and satisfy $\Lambdab_i\iN\pk$, 
and analyze the tensor product space (over \Complex)
  \be  \hvl:=\hil_{\Lambda_1} \otimes \hil_{\Lambda_2}
  \otimes\cdots\otimes\hil_{\Lambda_m}  \,.  \labl{hvl}
This tensor product is in a natural way a module over the $\mn$-fold direct sum
$\gM^m\equiv\gM\oplus\gM\oplus\cdots\oplus\gM$.

In addition to \ihwm s, we will occasionally also have to deal with
other objects in the category $\calo$, namely with the
so-called {\em generalized\/} or {\em parabolic Verma modules\/}
  \be  \VV_\Lambda:= \U(\gM) \otimes^{}_{\U(\pM)} \hlb  \,. \Labl VV
Here $\hlb$ denotes the irreducible \gb-module with \hw\
$\Lambdab$, while $\pM$ is an arbitrary parabolic subalgebra of \g.
More specifically, we consider the case where
  \be  \pM=\gB \oplus \complex K \oplus \gplus \,.  \labl{pM}
Then $\gM\eq\pM\opluS\gminus$ with $\gminus\eq\gB\otimes t^{-1}\complex[t^{-1}]$
as defined in \erf{gpm}, and for any integrable
\g-weight $\Lambda$ the \pbw\ implies a natural isomorphism
  \be  \VV_\Lambda \equiv \U(\gM) \otimes^{}_{\U(\pM)} \hlb
  \,\cong\, \U(\gminus) \otimes \hlb  \labl{vuu}
of $\gminus$-modules. In particular, $\VV_\Lambda$ is free as a 
$\gminus$-module.

Finally another class of modules will play a r\^ole, which are not restricted 
and hence in particular not in the category $\calo$, but still integrable. 
These \g-modules, called {\em evaluation modules\/}, are \findim\ and of level 
zero; we will encounter them in subsection \ref{sB}.
  
For later reference we also present the characters of some of the \g-modules of 
our interest. The character of an integrable \ihwm\ \hl\ is given by the \wkcf
  \be  \chila = \uvc \cdot \sumww \epsw\,\eE^{w(\Lambda+\rho)} \,,  \Labl wk
where the summation is over the Weyl group $W$ of \g, $\rho$ is the Weyl vector 
of \g\ and
  \be  \uvc:= \llb \sumww \epsw\, \eE^{w(\rho)}_{} \lrb^{-1}_{}
  = \eE^{-\rho} \prod_{\alpha>0}(1-\eE^{-\alpha})^{-{\rm mult}\,\alpha}_{}
  \,.  \labl{uvc}
Both $\uvc$ and the second factor in \Erf wk are totally antisymmetric under
the Weyl group $W$, so that $\chila$ is $W$-invariant.
The character of the Verma module with \hw\ $\Lambda$ is
$\eE^{\Lambda+\rho}\uvc$. Accordingly we will refer to the quantity \erf{uvc} 
as the `universal Verma character' of \g.

\subsection{The tensor product \HVL\ as a \gcP-module}\label{HVL}

We would now like to endow the vector space \erf{hvl} with the structure 
of a \gcP-module. To this end we have to choose a local holomorphic
coordinate $\z_i$ around each insertion point $p_i$
such that $\z_i(p_i)\eq0$; for instance, in terms of the global coordinate 
$z$ of $\complex\subseT\pe$, we can take $\z_i=z\mi z_i$ when $p_i\nE\infty$, 
while for $p_i\eq\infty$ we can take $\z_i=z^{-1}$. 

For any $\xb\otim f$ with $\xb\iN\gB$ and $f\iN{\cal F}(\Pem)$ we expand $f$ in
these local coordinates 
so as to obtain Laurent series $f_{p_i}\eq\fpi(\z_i)$. By linear extension
this defines, for each $i\inonetom$, a ring homomorphism from ${\cal F}(\Pem)$
to $\complex\twobrac{\z_i}$, and hence by identifying the indeterminate $t$ of
the loop construction with the local coordinate $\z_i$, the local realizations
  \be  x_i:=\xb\oT f_{p_i}  \labl{xi}
can be regarded as elements of the loop algebra
  \be  \gl = \gB\otimes\complex\twobrac t = \gminus \opluS \gB \opluS \gplus
  \,,  \ee
and, as such, as elements of the affine \lie\ \g\ \erf{gM}. Doing so, along with
any vector $v_i\iN\hli$ also $R_{\Lambda_i}(x_i) v_i$, where $R_{\Lambda_i}$ is
the \g-\rep\ associated to the irreducible \g-module $\hli$, is a vector in 
\hli. Moreover, even though the \bla\ \gcP\ is not centrally extended, 
we can obtain a \rep\ \rvl\ of \gcP\ on \hvl, namely by defining
the action of $\xb\otim f\iN\gcP$
on the element $v_1\otim v_2 \otim\cdots\otim v_m\iN\hvl$ by
  \be   \llb \rvl(\xb\otim f)\lrb (v_1\otim v_2\otim\cdots\otim v_m)
  := \sum_{i=1}^m v_1\otim v_2 \otim\cdots\otim\, R_{\Lambda_i}(x_i) v_i\,\otim
  \cdots\otim v_m \,.  \labl{rxf}
To verify that this yields a \gcP-\rep\ we compute
  \be  \hsp{-2}\bearl
  \llb \rvl(\xb\otim f)\rvl(\yb\otim g)\lrb (v_1\otim v_2\otim\cdots\otim v_m)-
  \llb \rvl(\yb\otim g)\rvl(\xb\otim f)\lrb (v_1\otim v_2\otim\cdots\otim v_m)
  \\{}\\[-.7em] \hsp{12}
  = \summi v_1\otim v_2 \otim\cdots\oT
    [R_{\Lambda_i}(x_i),R_{\Lambda_i}(y_i)] v_i\oT \cdots\otim v_m
  \\{}\\[-.7em] \hsp{12}
  = \summi v_1\otim v_2 \ot\cdots\oT R_{\Lambda_i}([\xb,\yb]\ot f_{p_i}g_{p_i})
    v_i\oT \cdots\otim v_m \\[.2em] \hsp{13.6}
    + \kv\,\kappa(\xb,\yb)\, \llb \summi \resi({\rm d}f\,g) \lrb
    \, v_1\otim v_2 \otim\cdots\otim v_m \,, \eear \labl{xfyg}
where in the first equality we observed that terms acting on different tensor
factors of \hvl\ cancel and in the second equality we inserted
the bracket relations of the affine \lie\ \g.
Now the terms in \erf{xfyg} that involve the level \kv\ cancel as a
consequence of the residue formula, while the other terms add up to
$\rvl([\xb\otim f, \yb\otim g])(v_1\otim v_2 \otim\cdots\otim v_m)$, where the
Lie bracket is the one of \gcP\ as defined in \Erf xy. Hence as promised,
for any choice $\vec\zeta\,{\equiv}\,(\zeta_1,\zeta_2\Ldots \zeta_m)$ of local 
coordinates at the parabolic points we have a \rep\ $\rvl=\rvlz$ of the block 
algebra \gcP.

Note that the cancellation of the terms coming from the central extension
only works if all modules \hli\ have the same level. Under the same condition
the result does not only hold for \ihwm s, but analogously also for any module 
on which the central element $K$ acts as a multiple of the identity,
and hence in particular for \vermod s and their quotients and for direct sums
of such modules. 

\subsection{\Coin s}

Before we can introduce chiral blocks, we need one more ingredient, the notion
of a \coin. For any \lie\ $\h$ we denote by $\U(\h)$ its universal enveloping 
algebra and by
  \be  \Up(\h):=\h\,\U(\h)  \ee
the augmentation ideal of $\U(\h)$. Then for any $\h$-module $V$, the quotient
module
  \be  \coi V\h := V \,/\, \Up(\h)\,V   \labl,
is known as the {\em space of \coin s\/} of $V$ \wrt $\h$.
(Strictly speaking, in place of \erf, one should write 
  \be  \coi V\h := V \,/\, R(\Up(\h))\,V  \,,  \ee
where $R$ denotes the \rep\ by which $\h$ acts on the module $V$. Different 
actions of $\h$ on one and the same underlying vector space will of course
give rise to different spaces of \coin s.)

Let us list briefly a few basic facts about \coin s (for further properties
of \coin s see Appendix \ref{acoin}).\\
$\bullet$\,~The vector space $\coi V\h$ can be characterized as the largest 
quotient module of $V$ on which $\h$ acts trivially. \\
$\bullet$\,~The space of \coin s of the tensor product $V \otimeS W$ of two 
$\h$-modules $V$ and $W$ equals their tensor product over $\U(\h)$:
  \be  \coi{V\otimeS W}\h\,{\equiv}\,\coi{V\,{\otimesc}\,W}\h = V\,{\otimes^{}
  _{\U(\h)}}\,W \,.  \ee
Indeed, both spaces are by definition equal to the quotient of
$V \otimeS W$ by the subspace that is spanned by the vectors $(xv)\otim w+
v\otim(xw)$ with $v\iN V$, $w\iN W$ and $x\iN\h$.\,%
 \futnote{In the definition of the tensor product $\otimes^{}_{\U(\h)}$ we
have to include the canonical anti-involution of $\U(\h)$ that is 
defined by $\bfe\mapsto\bfe$ and $x\mapsto-x$ for all $x\iN\h$, so as
to obtain the structure of a right $\U(\h)$-module on $V$.}
\\
$\bullet$\,~The concept of \coin s generalizes the notion of
invariant tensors to the case of non-fully reducible modules. More precisely, 
when $\h$ is a \findim\ semi-simple \lie\ and $V$ is fully reducible, then
$\coi V\h$ is isomorphic to the space of invariant tensors of $V$. Thus
the dimension of $\coi V\h$ is given by the number of singlets
contained in $V$; in particular, one has the formula
  \be  \dim \llb \coi{\hlb \otimeS \hmb}\h\lrb
  = \delta^{}_{\Lambdab',\Lambdab\op}  \Labl0f
for the $\h$-\coin s of the tensor product of two \findim\ \hwm s over $\h$;
a distinguished representative for the corresponding non-trivial \coin\ is 
given by $v_{\bar\Lambda}\ot v_{-(\bar\Lambda')^+}$, where $v_\Lambda$ is the 
highest weight vector of $\hlb$ and $v_{-(\bar\Lambda')^+}$ the lowest 
weight vector of $\hmb$.

\subsection{The definition of chiral blocks}\label{ss.d}

We are now in a position to give a precise definition of the chiral blocks
\cite{tsuy,falt,beau,beLa,fesv2}. We have seen that any choice $\vec\zeta\,
{\equiv}\,(\zeta_1,\zeta_2\Ldots \zeta_m)$ of local coordinates at the parabolic
points leads to a representation \rvlz\ of the block algebra $\gcP$ on \hvl. 
We start by introducing the spaces
  \be  \B_{\vec\zeta} = \coi\hvl{\Rvlz(\gcPM)} \equiv \hvl \,/\, 
  \rvlz(\Up(\gcP))\, \hvl \,  \Labl B'
of \coin s. Now these spaces of course depend on the choice of local 
coordinates $\vec\zeta$.
On the other hand, as already mentioned the chiral blocks play the r\^ole of
the building blocks for correlation functions in \cft; it is a fundamental
physical requirement that those correlation functions
should depend covariantly on the choice of coordinates. Here we will impose the 
natural stronger requirement that even the chiral blocks transform covariantly
under a change of the local coordinates at the parabolic points.

To get rid of the coordinate dependence we make use of the group
  \be U := \{u\iN\complex[[z]] \,|\, u(0)\eq0,\, \Frac{{\rm d}u}{{\rm d}z}(0)\nE0
  \} \ee
of local coordinate changes. The group $U^m\equiv U\Times U\Times\cdots\Times U$ 
acts transitively on the set of all collections $\vec\zeta$ of local
coordinates by sending each local coordinate $\zeta_i$ to the local
coordinate $\zeta_i\Circ u_i$ with $u_i\iN U$. Moreover, for any $\vec u\iN U^m$
one can find a map $\vGu{:}\ \hvl\,{\to}\,\hvl$ with the property that 
  \be  \vGu\, \rvlz(x) = \rvlu(x)\, \vGu \ee
for all $x\iN\gcP$, and this map is unique up to a scalar. 
Now via the local realizations \Erf xi (obtained by identifying the local 
coordinates at the parabolic points with the indeterminate of the loop 
construction for the relevant summand of $\gM^m\equiv\gM\oplus\gM\oplus\cdots
\oplus\gM$) one associates to any choice $\vec\zeta$ of local coordinates a 
sub\alg\ $\gmz$ of $\gM^m$ 
which is isomorphic to the \bla\ \gcP. The map $\vGu$
has the property that it restricts to an isomorphism
  \be  \vGu:\quad \gmz \hvl\stackrel\cong\to \gmzu \hvl  \ee
and therefore induces a map on spaces of \coin s taken with respect
to the different actions of the block algebra on the space \hvl\
that are associated to the different choices of local coordinates.

We now interpret the chiral blocks as the {\em equivalence classes\/} of \coin s
under the action of the group $U^m$, and denote the blocks by
  \be  \B = \coi\hvl\gcPM \equiv \hvl \,/\, \Up(\gcP)\, \hvl \,,  \labl B
where we think of $\gcP$ as the abstract \bla\ $\gB\otimeS{\cal F}(\Pem)$
without specifying its embedding into  $\g^m$.
For the purposes of this paper, we regard this prescription as the definition
of the spaces $\B \equiv \B^{}_{\!\{\Lambda_i\}, \{p_i\}}$  of chiral blocks. 

In \cft\ terminology, taking \gcP-\coin s corresponds to the procedure of
imposing the Ward identities of the current \alg\ on the correlation functions.
Let us also remark that in certain contexts it is actually more natural to 
consider the dual of the space $\coi\hvl\gcPM$, i.e.\ the space of 
$\gcPM$-invariants in the (algebraic) dual of $\hvl$. As already mentioned, in 
algebraic geometry and in quantum Chern\hy Simons theory, \cb s describe 
holomorphic sections in line bundles over moduli spaces; the relation to 
invariants is via \infdim\ Borel\hy Weil\hy Bott theory in which the full 
algebraic dual of a highest weight module appears. Here we are only interested
in the dimensions, which are finite and hence are the same for invariants and
\coin s.
 
Manifestly, the definition \erf B does not depend on the choice of local
coordinates. Moreover, any two spaces of chiral blocks with the same number of
insertion points and the same highest weights are isomorphic.
(The relevant isomorphism is canonical when there is a conformal mapping
of \pe\ that maps the two sets of insertion points bijectively on each other.)
Later on we will, however, often work with specific representatives, i.e.\
prescribe a specific choice of coordinates.  In this paper, we will focus on a 
genus zero curve for which a (quasi-)\,global holomorphic
coordinate exists. Any choice of such a global coordinate gives a natural
set of local coordinates at the parabolic points.

\Sect{Two-point blocks}T

The Verlinde formula provides a closed expression for the dimensions $\dim\B$
of spaces of chiral blocks. To enter the calculation of such dimensions,
let us first investigate the special situation where the number of insertion 
points is $m\eq2$. Recall that
by definition the chiral blocks are independent of the choice of the
global holomorphic coordinate $z$. For definiteness, we will work with specific
representatives of the blocks, namely by choosing the global coordinate in
such a way that the two insertion points are at $z\eq0$ and at $z\eq\infty$.
Then the block algebra is nothing but the (polynomial) loop algebra
  \be  \gbar(\pe{\setminus}\{0{,}\infty\}) = \gB\otimes\complex[z{,}z^{-1}]
  =: \gt \,, \Labl gt
where the indeterminate of the loop construction is given by the global 
coordinate $z$ on $\complex\eq\pe{\setminus}\{\infty\}$.
Now as we have seen in Subsection \ref{HVL}, the tensor product 
$\hvz\equiv\HL1\otimES\HL2$ is a \gT-module. In order to describe 
the \gT-\coin s of this module it is, however, most desirable that not only
the tensor product, but both factors $\HL1$ and $\HL2$
can be regarded as modules over the block \alg\ individually.
As is clear from the calculation \erf{xfyg}, this can definitely not be
achieved with the block \alg\ as defined so far. Therefore in addition to \gT\
we introduce a central extension of \gT\ by a one-dimensional center 
$\complex\Kt$, so as to obtain a centrally extended loop algebra 
$\gtc=\gt\oplus\complex\Kt$. The bracket relations of $\gtc$ read
  \be  [\xb\ot f,\bar y\ot g]=[\xb,\yb]\oT fg + \kappa(\xb,\yb)\,
  \Res^{}_0({\rm d}f\,g) \,\Kt \,.  \ee

By identifying the local coordinates at $0$ and $\infty$, \resp, with the
indeterminate of the loop construction, we obtain two different embeddings 
$\j_0$ and $\j_\infty$ of $\gt$ as a vector space into \g. We can
extend these maps to two isomorphisms between  $\gtc$ and \g\ as \lie s, 
which read explicitly
  \be  \j_0(\xb\otim z^n):= \xn\,, \hsp{2.9} \j_0(\Kt):= K  \hsp{1.6}  \labl8
($\xb\iN\gb$, $n\iN\zet$) and
  \be  \j_\infty(\xb\otim z^n):= \xmn\,, \qquad \j_\infty(\Kt):= -K \,,  \labl9
\resp. (Note that one should carefully distinguish between the \bla\ as an 
abstract \lie\ and the embedding of the \bla\ into $\g^m$ via local coordinates.
For instance, in the two-point situation considered here, the \bla\ 
\gT\ is isomorphic to the \alg\ $\gB\otimeS\complex(t)$ of \gb-valued
Laurent {\em polynomials\/}
that is contained in $\oog$. But for a generic choice of the coordinate $z$,
\gT\ is embedded into \g\ as a sub\alg\ that is isomorphic but not identical 
to $\gB\otimeS \complex(t)$ and in particular involves arbitrary Laurent series,
as is generically needed in order to be able to define inverses and hence to
have a {\em group\/} of local coordinate changes. This is the main motivation 
why in this paper we regard \g\ rather than its sub\alg\ $\oog$ as the affine 
\lie.)\,%
 \futnote{As a side remark, we mention that we can extend the automorphism 
$\omega:=\j_0\circ(\j_\infty)^{-1} =\j_\infty\circ(\j_0)^{-1}$, which acts as
$\omega(\xn)\eq\xmn$ for $\xb\iN\gb$ and $n\iN\zet\,$ and as 
$\,\omega(K)\eq{-}K$, to an automorphism that includes the outer derivation on 
the centrally extended loop \alg, namely via $\,\omega(D)\df{-}D$.}

Using the embeddings \erf8 and \erf9, each of the two irreducible highest 
weight modules $\HL1$ and $\HL2$ of \g\ can be 
separately endowed with the structure of a $\gtc$-module. With the help of
the isomorphism $\j_0$ we can transport the triangular decomposition \Erf1t
of \g\ to a triangular decomposition of $\gtc$. (Alternatively we could
choose to use $\j_\infty$ for this purpose; this would result in a different
triangular decomposition.) It then makes sense to talk about highest and lowest 
weight modules over the centrally extended \bla\ $\gtc$. 
Let us analyse the structure of the vector spaces $\HL1$ and $\HL2$
in this spirit. One easily sees that also as a $\gtc$-module, $\HL1$ is an 
irreducible highest weight module of level \kv\ with highest weight $\Lambda_1$,
while as a $\gtc$-module $\HL2$ is a {\em lowest\/} weight module, with lowest
weight $-\Lambda_2^+$ and at level $-\kv$.\,%
 \futnote{Thus the horizontal part of the weight is $-\Lambdab_2^+$, which is
the lowest weight of the \findim\ \gb-module with \hw\ $\Lambdab_2$ or, in
other words, $-\Lambdab_2^+\eq\wmax(\Lambdab_2)$ with $\wmax$ the longest
element of the Weyl group of \gb.
Also, even though $\HL2$ as a $\gtc$-module is a lowest weight module
and hence in particular not in the category $\calo$, the distinction
between the \alg s \g\ and $\oog$ is again immaterial, because the module is 
integrable and restricted.}
We are interested in the tensor product of these two $\gtc$-modules, and in the
\coin s of the tensor product \wrtt action of $\gt$. Now we make the simple 
but crucial observation that
the tensor product $\HL1\otimES\HL2$ has level $\,\kv+(-\kv)\eq0\,$ as a
$\gtc$-module, or in other words, that the $\gtc$-module $\HL1\otimES\HL2$ 
factorizes to a \gT-module. As a consequence, the \gT-\coin s of the tensor 
product {\em coincide\/} with its $\gtc$-\coin s.

We can conclude that we are left with the task of finding the $\gtc$-\coin s of
$\HL1\otimES\HL2$. Now it is not difficult at all to determine these \coin s 
explicitly. Namely, every vector in $\HL1\otimES\HL2$ is a finite sum of
vectors $v_1\ot v_2$ with $v_1\iN\HL1$ and $v_2\iN\HL2$, and as a consequence
without loss of generality we can take such vectors $v_1\ot v_2$ as
representatives for the $\gtc$-\coin s. Now since $\HL1$ is a \hwm\ over $\gtc$,
we can write $v_1\eq x_-v_+$, where $v_+$ is a highest weight vector of $\HL1$ 
and $x_-\iN\U(\gtc_-)$ with $\gtc_- \equiv \j_0^{-1}(\g_-)$.
Moreover, as representatives of $\gtc$-\coin s, we have the equivalence
  \be v_1\oT v_2 = x_- v_+ \oT v_2 \;\sim\; v_+ \oT y_- v_2 \,,  \ee
where $y_-\eq (-1)^n\,x_n x_{n-1}\cdots x_2 x_1$ when $x_-\eq x_1 x_2\cdots x_n$
with $x_i\iN\gtc_-$.
Without loss of generality we can therefore assume that $v_1\eq v_+$, i.e.\
restrict our attention to representative vectors of the form $v_+ \ot v_2$.
Next we write $v_2\eq y_+ v_-$, where $y_+\iN\U(\gtc_+)$ and $v_-$ is the lowest
weight vector
of $\HL2$, regarded as a $\gtc$-module. Since any part of $y_+$ that is in the
augmentation ideal of $\U(\gtc_+)$ annihilates $v_+$, we can use the same
argument as before to conclude that the only possible representative for a
$\gtc$-\coin\ is $v_+\ot v_-$. Finally we impose invariance under the \csa\
of $\gtc$, which amounts to the requirement that $v_+\ot v_-$ has weight zero, 
i.e.\ that
  \be  \Lambda_1+(-\Lambda_2^+)=0 \,.  \ee
Thus we conclude that the space of \coin s is zero-dimensional unless 
$\Lambda_1\eq\Lambda_2^+$, in which case it is \onedim\ and has $v_+\otim v_-$ 
as a (distinguished) representative. This result (and its derivation, too) is in
complete analogy with the formula \Erf0f for \coin s of \findim\ semi-simple 
\lie s.

In short, we have shown that
  \be  \dim\B_{\{\Lambda_1,\Lambda_2\}} = \delta^{}_{\Lambda_1^{},\Lambda_2^+}
  \,.  \Labl2d
Of course, this simple result can also be obtained by various other means.
The reason why we presented this particular derivation is that it sets the
stage for a similar analysis that can be performed for any arbitrary value $\mn$
of insertion points.

\Sect{An integral formula for $\dim B$}A

\subsection{The space \B\ as a two-point \coin}\label{sBH}

Let us now turn to the $m$-point situation, where $m>2$. 
Our aim is to describe the space
  \be  \B= \coi\hvl\gcPM \equiv \coi{\, \bigotimes_{i=1}^\mn \hli}\gcPM
  \labl{BH}
of chiral blocks. Because of the independence of \B\ of the choice of 
coordinates, we can assume that the first and $\mn$th insertion points $p_1$ 
and $p_\mn$ are at $z_1\eq0$ and $z_\mn\eq\infty$, \resp.
It can then be shown that there is a natural isomorphism
  \be  \B \;\cong\; \coi{\HL1 \otimeS \barB \otimeS \HL\mn}\gt  \labl{biso}
of vector spaces, where \gT\ is the block \alg\ \Erf gt corresponding to only
two insertion points at zero and infinity,
and where $\barB$ stands for the tensor product
  \be \barB := \bigotimes_{i=2}^\mm \hlib   \ee
of \findim\ irreducible \gb-modules.
Here \hlib\ denotes the \ihwm\ of \gb\ whose \hw\ $\Lambdab_i$ is the horizontal
projection of the \hw\ $\Lambda_i$ of the \g-module \hli. Furthermore,
\gT\ is defined to act on $\barB$ by evaluation in the obvious manner, 
i.e.\ $\xb\ot f\iN\gt$ with $\xb\iN\gB$ acts as 
$\sum_{i=2}^\mm f_{p_i}(0)\cdot \bfe\oT \cdots \oT \bfe\oT
  \bar R_{\Lambdab_i}(\xb)\oT\bfe \oT \cdots \oT \bfe $,
where $\bar R_{\Lambdab_i}$ denotes the \gb-\rep\ carried by \hlib\
and $f_{p_i}(0)\eq f(z_i)$ is the value of $f\iN{\cal F}(\Pem)$ at the
insertion point.\,%
 \futnote{Note that the result \erf{biso} implies in particular that \coin s 
taken with respect to different actions of the \bla, corresponding to
different sets of local coordinates, are isomorphic. Namely, after replacing
the relevant affine module by the corresponding \findim\ module the only data
entering the description of the block are the values of the functions 
$f\iN{\cal F}(\Pem)$ at the insertion point; manifestly these values do not
depend on the choice of local coordinates.}

The proof of the isomorphism \erf{biso} between the space of
chiral blocks and the space $\coi{\HL1\otimeS\barB\otimeS\HL\mn^{}}\gt$
of \gT-\coin s is not difficult, but a bit lengthy
(compare also proposition 2.3.\ of \cite{beau}).
Therefore we present here only an outline of the proof and defer most details
to Appendix \ref{aproof}.

Rather than proving the isomorphism \erf{biso} directly, we start from a
somewhat more general setting. We consider two finite sets
$P:=\{p_1,p_2\Ldots p_n \}$ and $Q:=\{q_1,q_2\Ldots q_\lm\}$ of pairwise
distinct points of $\pe$, where we assume that the set $Q$ is not empty,
while $P$ may be empty. To each point we 
associate an integrable highest weight of \g, and introduce the tensor product
  \be  \neh := \nehe \otimes \nehz  \qquad{\rm with}\qquad
  \nehe := \bigotimes_{i=1}^n \hilb_{\Lambdab_i} \, ,  \quad
  \nehz := \bigotimes_{j=1}^\lm \hil_{\Lambda'_j}  \Labl ht
of irreducible modules of the horizontal subalgebra \gb\ for the points in $P$
and of irreducible modules of the affine \lie\ \g\ for the points in $Q$, \resp.
Finally, we fix an additional insertion point on $\pe\,{\setminus}(P\cup Q)$,
with an associated integrable highest \g-weight $\Lambda$. Without loss of 
generality we can assume that this insertion point is $z=\infty$.

Moreover, in addition to \hl\ we also consider the corresponding parabolic Verma
module
$\Verl=\U(\g)\,{\otimes_{\U(\pM)}}\,\hlb \,{\cong}\, \U(\gminus)\otimeS\hlb$
as introduced in formula \erf{vuu}. As we will show in Appendix \ref{aproof} 
by employing special properties of \coin s of free modules and the
behavior of exact sequences under the operation of taking \coin s,
we have the isomorphisms
  \be  \coi{\neh\otimes\hilb_\Lambdab}{\gB(\pE\setminus Q)} \;\cong\;
  {\lfloor \coi{\neh\otimes \VV_{\Lambda}}\gminus \rfloor}_{\gB(\pE\setminus Q)}
  \;\cong\; \coi{\neh\otimes\VV_{\Lambda}}{\gB(\pE\setminus (Q\cup\{\infty\}))}
  \labl{ihV}
and
  \be  \coi{\neh\otimeS\Verl}{\gB(\pE\setminus(Q\cup\{\infty\}))} \;\cong\;
  \coi{\neh\otimeS\hl}{\gB(\pE\setminus(Q\cup\{\infty\}))} \,.  \labl{hV}
Combining these results, we learn that there is an isomorphism
  \be  \coi{\neh\otimeS\hl}{\gB(\pE\setminus(Q\cup\{\infty\}))} \;\cong\;
  \coi{\neh\otimes\hilb_{\Lambdab}}{\gB(\pE\setminus Q)} \,.  \labl{47}
Employing this isomorphism we replace successively all but the first and last\,%
 \futnote{If we so wished, we could also replace one of the remaining \g-modules
$\HL1$ and $\HL\mn$ by the corresponding \gb-module, thereby obtaining \coin s 
\wrtt one-point block \alg\ $\gB(\pe{\setminus}\{\infty\})\eq\gB(\complex)\eq\gB
\otimeS\complex[[z]]\,{\cong}\,\gB\opluS\gplus$. In contrast, according to the
discussion in Appendix \ref{aproof}, it is not possible to apply this
manipulation also to {\em both\/} of the remaining \g-modules.}
affine \irmod s \hli\ that appear in the space \erf{BH} of \coin s by
the corresponding \irmod s \hlib\ of the horizontal subalgebra \gb.\,%
This way one finally arrives at the isomorphism \erf{biso}.

\subsection{Branching rules}\label{sB}

Now we realize that
in view of the isomorphism \erf{biso} we are actually in a situation very
similar to the one investigated in \sref T. The extension to the general case 
amounts to take properly into account the additional \findim\ \gb-modules.
We will argue that this can be achieved by means of suitable branching rules
for completed tensor products involving \findim\ $\gtc$-modules.
First we note that all elements of $\gt$ are represented 
on the tensor product $\barB$ of \findim\ \gb-modules by evaluation at the
respective insertion points $p_2,p_3\Ldots p_{m-1}$, so that
no central term can arise in their Lie brackets. Therefore the action of 
\gT\ on $\barB$ can be extended to an action of $\gtc$ with $\Kt$ represented by
zero. We call the thus obtained \findim\ level-zero $\gtc$-module $\barB$,
as well as any other $\gtc$-module that is obtained in an analogous manner
from a \findim\ \gb-module, an {\em evaluation module\/}. 
Evaluation modules are not restricted, hence not in the category $\calo$,
but the step operators of $\gtc$ still act locally nilpotently.

As a consequence, the tensor product $\HL1\otimeS\barB$ of $\gtc$-modules 
isn't an object in the category $\calo$ either; but
it is easily checked that this $\gtc$-module is still integrable (this fact is
also used in the proof of \Erf hV in Appendix \ref{aproof}),
and that the central element $\Kt$ acts as multiplication by \kv.
Not surprisingly, it turns out to be rather difficult to compute the \coin s 
of the module $\HL1\otimES\barB$ directly.
In order to determine the \coin s, we would therefore like to replace 
$\HL1\otimES\barB$ by a direct sum of \ihwm s of $\gtc$.
To this end we work with a completion \hhc\ of $\HL1\otimES\barB$.
More specifically, we assume that there exists a suitable a completion such 
that \hhc\ becomes reducible and can be written as a direct sum
  \be  \hhc\ \,\cong\, \rest \oplus \bigoplus_\ell\HM\ell \,,  \labl{deco}
where each of the summands $\HM\ell$ is an integrable highest weight 
$\gtc$-module, while $\rest$ is a direct sum of
$\gtc$-modules which are irreducible, but are not weight modules. 
(Note that for the uncompleted module such a decomposition 
typically does not exist.)

Now the tensor product $(\bigoplus_\ell\HM\ell)\otimes\HL\mn
\subseteq\hhc\otimeS\HL\mn$ is at level $\kv+(-\kv)=0$\, as a $\gtc$-module,
and hence by repeating the arguments of \sref T one finds that its \gT-\coin s
are the same as its $\gtc$-\coin s. The $\gtc$-\coin s in turn can be
computed in exactly the same manner as in the two-point situation of
\sref T. More precisely, among the \inthwm s in the decomposition \erf{deco}
only those modules contribute to the $\gtc$-\coin s whose highest weight is 
$\Lambda_\mn^+$, and each of these yields precisely one independent \coin. 
In addition, the same analysis indicates 
that the submodule $\rest$ of \hhc\ does not contribute any \coin s, and also 
that the $\gtc$-\coin s of $(\hh)\otimeS\HL\mn$ are the same as those of
$\hhc\otimeS\HL\mn$.

Unfortunately, so far we cannot describe a topology on the module
$\calh_{\Lambda_1}\otimes\bar\calh^{(m-2)}$ such that in the 
completion \wrt that topology the relation \erf{deco} holds. The existence 
of such a topology should therefore be regarded as a conjecture.
{}In the sequel we will assume that this conjecture is valid; then
the computation of the dimension $\dim\B$ of the space \Erf BH of \coin s
merely amounts to determining the \inthwm s in the decomposition \erf{deco} of
$\HL1\otimES\barB$. Rewriting the branching rule \erf{deco} as
  \be  \HL1\otimes\barB \,\cong\, \rest \oplus
  \bigoplus_{\mu\in\pk} \NN\mu\,\hil_\mu \,, \labl-
our arguments amount to the formula
  \be  \dim\B_{\!\{\Lambda_i\},\{p_i\}} = \NNL \,.  \labl+

Notice that on the module $\calh_{\Lambda_1}\otimeS\bar\calh^{(m-2)}$
(as on any other evaluation module of the affine \lie, except for the
trivial module) we do not have an action of the derivation in the \csa\
of the affine \lie\ \g. However, a posteriori, we can define such an
action on a submodule of the completion of 
$\calh_{\Lambda_1}\otimeS\bar\calh^{(m-2)}$, namely on the direct sum of the
\inthwm s $\hil_\mu$ over the centrally extended block \alg\ $\gtc$. We will
assume from now on that this has been done and that the eigenvalues of the
derivation on the \hwv s are chosen in the standard way such that 
the characters furnish a module of \slz.
This, together with the assumption that the module $\rest$ is not a weight
module and hence does not contribute to characters, opens the possibility to 
compute the branching coefficients $\NN\mu$ by manipulating characters, and 
indeed this will be achieved in the next subsections. Notice that the
character of $\bigoplus_{\mu\in\pk} \NN\mu\,\hil_\mu\subseteq\hhc$ as 
introduced in formula \erf- is given by the product of the character $\chil1$ 
of $\HL1$ and the \gb-characters $\chibl2$ to $\chibl{\mn-1}$,
   \be  \chii^{}_{\oplus^{}_\mu\NN\mu\hiL_\mu} = \chil1(h)\cdot \chibarX(\hb)
   \,,  \Labl0c
where it is understood
that the eigenvalues of the derivation are chosen as explained above.

\subsection{A projection formula}

Our task is now to find the \inthwm s in 
the decomposition \erf{deco}. In this subsection 
we present a projection formula which can be used to that effect. We start by
introducing a certain linear functional $\PL$ on the space of class functions 
$\varphi$ on $G$, where $G$ is the connected, simply connected 
compact real Lie group whose \lie\ is the compact real form
of the horizontal subalgebra \gbar\ of \g. We denote by $T$ the maximal
torus of $G$ that corresponds to the chosen \csa\ \gbo\ of \gb. Thus the 
elements $t$ of $T$ are group elements of the form 
  \be  t(\hb) = \Exp(\zpi\hb)  \quad{\rm with}\;\ \hb\iN(\gbo)_\reals^{} \,,
  \ee
and the normalized Haar measure $\dt$ on $T$ is the flat measure
$\intT\eq\prod_{i=1}^{{\rm rank}\,\gB}(\int_0^1\!\rmd\hb^i)$.

We then define, for each integrable weight $\Lambda$ of \g, the linear 
functional \PL\ by
  \be  \PL[\varphi] := \intT \llb \uvc^{-1} \eE^{-(\Lambda+\rho)}\lrb(\zpi h)
  \cdot \varphi(\hb)  \,.  \ee
Here $\uvc$ is the `universal Verma character' that appears in the \wkcf\ 
\Erf wk. Also, concerning the relation between elements of \go\ and of \gbo\ 
and between elements of their dual spaces we adhere to the following 
conventions. The elements $\hb\iN\gbo$ and $h\iN\go$ are related by
  \be  h = \hb - \tau L_0 + \zea K  \ee
with $\tau,\zea\iN\complex$, and the analogous decomposition of
\g-weights reads
  \be  \mu = \mub+\mu^0\lo+\nnu\delta =: \mub + \mu\afp \,,  \ee
where $\lo$ is the zeroth fundamental \g-weight and $\delta$ the null root of 
\g. Thus the \g-weight $\mu\iN\godual$ acts on $h\iN\go$ as
  \be  \mu(h)=\mub(\hb)+\mu\afp(\tlk)=\mub(\hb)+\zea\mu^0+\nnu\tau  \,.  \ee
Notice that strictly speaking, $\varphi$ has to be a functional on the full
\csa\ of the affine \lie\ \g, including the derivation; we will
not always explicitly display this dependence in the sequel.

Let us now consider the special case that $\varphi$ is the character of
a Verma module, $\varphi\equiv\varphi(\hb{;}\tau{,}\zea)
=\uvc_{\Lambda'}(\zpi h)$; we get immediately
  \be \bearll  \PL[\uvc_{\Lambda'}] \!\!
  &=\dstyle\intT \llb \uvc^{-1} \eE^{-(\Lambda+\rho)}\lrb(\zpi h) \cdot
    \llb \uvc \eE^{\Lambda'+\rho}\lrb(\zpi h)  \\{}\\[-.8em]
  &=\dstyle\intT \eE^{\Lambda'-\Lambda}(\zpi h)  \\{}\\[-.8em]
  &=\eE^{\zpi(\Lambda'-\Lambda)\afp(\tlk)}_{}\,\delta_{\bar\Lambda,\bar\Lambda'}
    \equiv \eE_{}^{\zpi\zea((\Lambda')^0-\Lambda^0)} \eE_{}^{\zpi\tau
    (\nu_{\Lambda'}-\nu_\Lambda)}\, \delta^{}_{\bar\Lambda,\bar\Lambda'}
  \,.  \eear \labl{vere}
As a consequence, the operator $\PL$ picks the $\Lambda$-isotypic 
component in the resolution of any module. In particular, we can analyze
the irreducible characters, i.e.\ take
$\varphi\eq\chii_\Lambda(\zpi h)$. By the \wkcf\ we know that
$\chii_\Lambda = \sumww \epsw\,\uvc_{w(\Lambda+\rho)-\rho}$,
and hence \erf{vere} tell us that
  \be \bearll \PL[\chii_{\Lambda'}] \!\!
  &=\dstyle\sumww \epsw\, \PL[\uvc_{w(\Lambda'+\rho)-\rho}]  \\{}\\[-.8em]
  &=\dstyle\sumww \epsw\,
    \eE^{\zpi[w(\Lambda'+\rho)-(\Lambda+\rho)]\afp(\tlk)}_{}\,
    \delta^{}_{\what(\Lambdab'+\rhob),\Lambdab+\rhob} \,.  \eear  \Labl0h
Here $\what$ denotes the induced affine action of $w\iN W$ on the weight space
\gbodual\ of the \hsa, i.e.\
the action of $w$ on the horizontal part $\mub$ of a \g-weight $\mu\iN\godual$
yields the horizontal part of the $w$-transformed \g-weight:
  \be  \what(\mub)\eq\wmb \,.  \Labl wh
In the following, for notational simplicity we suppress the hat-symbol and 
just write $w(\bar\mu)$ for the horizontal projection $\wmb$ of $w(\mu)$.

Next we restrict to the case of our interest, where both $\Lambda$ and
$\Lambda'$ are integrable weights, and
use the fact that the Weyl group $W$ of \g\ acts freely on the interior 
of the Weyl alcoves. This property of $W$ implies that 
in fact only a single Weyl group element, namely 
$w\eq\id$, can give a non-zero contribution to the sum; \asac\ \Erf0h reduces to
  \be  \PLP[\chila] = \eE^{\zpi(\Lambda-\Lambda')\afp(\tlk)}_{}\cdot
  \delta^{}_{\Lambdab,\Lambdab'}  \,.  \ee
Moreover, we are only interested in weights $\Lambda$ at some fixed value \kv\ 
of the level. Accordingly we now further
restrict our attention to functionals $\PLP$ 
for which $\Lambda'$ is at level \kv\ as well; then $\Lambdab\eq\Lambdab'$
implies that also $\Lambda^0\eq\Lambda'{}^0$.
Finally we require that the $\delta$-component of the weights $\Lambda$
and $\Lambda'$ is already specified by the horizontal part of the weights 
(e.g.\ that it is given by $-\Delta_\Lambda\eq{-}(\Lambdab,\Lambdab\pl2\rhob)
/4(\kv\pl\gv)$, the natural value in \cft; owing to the fact that
for all $\nnu\iN\complex$, $\hl$ and $\hil_{\Lambda+\nnu\delta}$ 
are isomorphic as \g-modules,
this does not result in any loss of generality). Then the equality of
$\Lambdab$ and $\Lambdab'$ also enforces equality of the $\delta$-components,
so it follows that in fact we not only have $\Lambdab\eq\Lambdab'$, but even
$\Lambda\eq\Lambda'$.

Summarizing, we have proven that the projection formula
  \be  \PLP[\chila] \equiv
  \intT \llb \uvc^{-1} \eE^{-(\Lambda'+\rho)}\,\chila\lrb (\zpi h)
  =\delta_{\Lambda,\Lambda'}^{}  \labl1
holds for all integrable \g-weights $\Lambda$ and $\Lambda'$ 
at fixed level whose $\delta$-components depend only on the horizontal part.
More generally, acting with the operator $\PL$ on the character of a direct
sum of integrable \hwm s provides us with the multiplicity of the module \hl\
in that sum.\,%
 \futnote{Incidentally, the identity \erf1 can also be used to derive an
integral formula for branching functions for embeddings of affine \lie s.
Such expressions for branching functions have also been obtained in 
\cite{haso,hwrh2}.}

Note that when we interpret the parameter $\tau$ as a complex variable (rather
than some formal indeterminate), in order for the character $\chila(\zpi h)$ to
be a convergent power series in $\exp(\zpi\tau)$, it must be required that 
$\tau$ has positive imaginary part. However, the result \erf1 tells us 
e.g.\ that the expression
  \be  \PLP[\chila] 
  = \dstyle \sumww \epsw\,\eE^{\zpi[w(\Lambda+\rho)-(\Lambda'+\rho)]\afp(\tlk)}
  _{} \intT \eE^{\zpi (w(\Lambdab'+\rhob)-(\Lambdab+\rhob))(\hb)}_{} \ee
is independent of $\tau$, so that in the present context
this restriction is in fact rather irrelevant.
Similarly, in various formul\ae\ that we will deal with below a priori the 
imaginary part of $\tau$ must be kept strictly positive in order that the 
affine characters, \resp\ the sums over the infinite Weyl group $W$, converge, 
and it is not guaranteed that the limit $\tau\,{\to}\,0$ of the whole expression
exists. 

\subsection{The integral formula}

We now employ the formula \erf1 to count the number of \inthwm s
with \hw\ $\Lambda_\mn^+$ in the completed tensor product \hhc. 
{}From our assumption \Erf0c about the characters we deduce
with the help of the projection formula \erf1 that the branching coefficients 
$\NN\mu$ appearing introduced in equation \erf- satisfy
  \be  \NNL = \intT \llb\uvc^{-1}\eE^{-\Lambda_\mn^+-\rho}\lrb(\zh) \cdot
  \chil1(\zh)\, \chibarX(\zhb)  \,.  \ee
We now manipulate the \rhs\ of this identity as follows.
We first insert the \wkcf\ \Erf wk for $\chil1$; next we introduce a dummy 
summation $|\Wb|^{-1}\sumwwb$ and substitute $w\mapstO\wbar^{-1}w$, where we
consider the Weyl group $\barW$ of \gb\ as canonically embedded in the Weyl 
group $W$ of \g; and finally we set $\hb\eq\wbar^{-1}(\hb')$ and use 
invariance of the measure $\dt$ under the transformation from $\hb$ to $\hb'$. 
This way we arrive at
  \begin{eqnarray}  & \NNL \!\!&
  = \dstyle \intT \chibarX(\zhb) \sumw w \eps(w)\,\eE^{w(\Lambda_{m-1}+\rho)
  -(\Lambda_\mn^++\rho)}(\zh)
  \nonumber\\{}\nonumber\\[-.3em]
  &&= \dstyle\Frac1{|\Wb|}\, \intT \chibarX(\zhb) \sumww \sumwwb
  \eps(w\wbar)\, \eE^{\wbar^{-1}w(\Lambda_1+\rho)-\Lambda_\mn^+-\rho}(\zh)
  \nonumber\\{}\nonumber\\[-.6em]
  &&= \dstyle\Frac1{|\Wb|}\, \intT \chibarX(\zhb)
  \nonumber\\{}\nonumber\\[-.98em]  &&\hsp{4.5}
  \cdot \dstyle\sumww \eps(w)\, \eE^{w(\Lambda_1+\rho)}(\zh) \cdot\sumwwb
  \eps(\wbar)\, \eE^{-\wbar(\Lambda_\mn^++\rho)}(\zh)
  \,. \end{eqnarray}
Next we use the Weyl character formula for the characters $\chilb$ of \findim\ 
\gb-modules \hlb\,%
 \futnote{In agreement with the remarks made above the final result, as 
displayed in formul\ae\ \Erf eN and \erf V below, does not depend
on $\tau$ and $\varpi$ at all. Accordingly, for ease of notation we suppress 
the superficial dependence on these variables from now on.}
to obtain
  \be \bearll  \NNL \!\!
  &= \dstyle\Frac1{|\Wb|}\, \intT \chibarX(\zhb)\,\chib_{\Lambdab^+_\mn}(-\zhb)
  \\{}\\[-.98em]&\hsp{11}
  \cdot \uvcb^{-1}(-\zhb)\,\dstyle\sumww \eps(w)\,\eE^{w(\Lambda_1+\rho)}(\zhb)
  \,, \eear \Labl*1
where 
  \be  \uvcb = \llb\! \sumwwb \epswb\, \eE^{\wbar(\rhob)}_{} \lrb^{-1}_{}
  = \eE^{-\rhob} \prod_{\alphab>0}(1-\eE^{-\alphab})^{-1}_{}  \ee
is the universal Verma character of \gb, the horizontal analogue of $\uvc$
\erf{uvc}.

In a final step we rewrite \Erf*1 in the form
  \begin{eqnarray}  && \phantom x\hsp{-2.5}
  \NNL = \dstyle\Frac1{|\Wb|} \intT \chibarX\chibl\mn(\zhb)
  \cdot \uvcb^{-1}(-\zhb)\! \sumww\! \eps(w)\, \eE^{w(\Lambda_1+\rho)}(\zhb)
  \,.\nonumber\\[-.6em] {}  \label*  \end{eqnarray}
Here we employed the identity
  \be  \bar\chii_{\Lambdab^+_{}}^{}(\hb) = \chilb(-\hb) \ee
for characters of \ihwm s of \gb. This relation follows by making the
substitution $\wbar\mapstO\wbar\wmax$, with $\wmax$ the longest element of the
Weyl group of \gb, in the \wcf, and by using the fact that this element acts as
$\wmax(\mub)\eq{-}\mub^+$ on \gb-weights:
  \be  \bar\chii_{\Lambdab^+_{}}^{}(\hb)
  = \frac {\sumwwb \epswb \eE^{\wbar(\Lambdab^++\rhob)(\hb)}}
  {\sumwwb \epswb \eE^{\wbar(\rhob)(\hb)}}
  = \frac {\eps(\wmax) \sumwwb \epswb \eE^{-\wbar(\Lambdab+\rhob)(\hb)}}
  {\eps(\wmax) \sumwwb \epswb \eE^{-\wbar(\rhob)(\hb)}} = \chilb(-\hb) \,.  \ee

\Sect{From the integral formula to the modular \smat}F

\subsection{$\dim\B$ in terms of weight multiplicities}\label{dbf}

Our aim is now to deduce the Verlinde formula from the integral formula \erf*. 
This can be achieved by using information about the structure of the Weyl 
group $W$ of \g\ or, more precisely, about the relation between $W$ and the
Weyl group $\barW$ of the \hsa\ \gb\ which is canonically embedded as a 
subgroup into $W$. There are essentially two different
possibilities to implement this relationship. The first amounts to
working with a special set $\Wp$ of representatives of the coset of 
$W$ by its subgroup $\barW$. By definition, $\Wp$ is 
that subset of $W$ which consists of those representatives of the elements of
$W/\barW$ that have minimal length. One knows
(see e.g.\ \cite{iwah2} and also \cite[remark\,8.1.]{gale}) that
every element of $W/\barW$ has a unique representative with this 
property, that for any integrable \g-weight $\Lambda$ and any $w\iN W$ the
\gb-weight $w(\Lambdab\pl\rhob)\mi\rhob$ is dominant integral if and only if
$w\iN\Wp$, and that each element $w\iN W$ can be uniquely represented in 
the form $w\eq\bar w \wp$ with $\wbar\iN\barW$ and $\wp\iN\Wp$.
Together with the \wcf, which absorbs the summation over $\barW$,
it then follows that our result \Erf*1 can be rewritten as
  \be \bearll  \dim\B \!\!
  &= \dstyle\Frac1{|\Wb|}\, \intT \chibarx \uvcb^{-1}(\zhb)
  \sumwwk\eps(\wbar\wkr)\,\chib_{\wkR(\Lambdab_1+\rhob)-\rhob}(\zhb)
  \\{}\\[-1.41em]&\hsp{20} \cdot\uvcb^{-1}(-\zhb)\,\chib_{\Lambdab^+_\mn}(-\zhb)
  \\{}\\[-.7em]
  &= \dstyle\Frac1{|\Wb|}\, \intT J(\zhb)\, \chibarX\chib_{\Lambdab_m}(\zhb)
  \sumwwk \eps(\wkr)\, \chib_{\wkR(\Lambdab_1+\rhob)-\rhob}(\zhb) 
  \,. \eear \Labl0d
Here we introduced the function
  \be  J(\zhb) := \uvcb^{-1}(\zhb)\, \uvcb^{-1}(-\zhb) = \prod_{\alphab\in
  \bar\Delta_+} \mbox{\Large$|$} \eE^{\pii\alphab(\hb)}-\eE^{-\pii\alphab(\hb)}
  \mbox{\Large$|$}^2 \,,  \labl{Jh}
where the product extends over the set $\bar\Delta_+$ of positive \gb-roots.
This function can be recognized as the Jacobian factor that, together with 
$|\barW|^{-1}$, appears in Weyl's integration formula that relates 
the integral of a class function
over the whole group $G$ to an integral over its maximal torus $T$.
Thus we can rewrite \Erf0d as an integral over $G$:
  \be  \dim\B = \sumwwk\eps(\wkr)\, \intG 
  \chib_{\wkR(\Lambdab_1+\rhob)-\rhob}\, \chibarX\chib_{\Lambdab_m}  \,, \ee
where $\dg$ is the normalized Haar measure of $G$. From the orthogonality
property of $G$-characters \wrt $\dg$ it then follows immediately that
  \be  \dim\B = \sumwwk \eps(\wkr)\,
  \bar N_{\!\wkR(\Lambdab_1+\rhob)-\rhob,\Lambdab_2...\Lambdab_\mn}  \,,
  \Labl eN
where $\bar N_{\!\mub_1\mub_2...\mub_l}$ denotes the number of singlets in
the tensor product $\hilb_{\mub_1}\otimeS\hilb_{\mub_2}{\otimes}\cdots\otimES
\hilb_{\mub_l}$.

The \rhs\ of expression \erf{eN} actually constitutes a finite sum, 
even though $\Wkr$ is an infinite set. To see this, we
observe that the \gb-module $\hilb_{\Lambdab_2}{\otimes}\cdots\otimES
\hilb_{\Lambdab_\mn}$ can be decomposed into a direct sum of finitely many  
irreducible \gb-modules. According to the identity \erf{0f},
the module $\hilb_{\wkR(\Lambdab_1+\rhob)-\rhob}$ 
has to appear in this decomposition in order to give a non-zero contribution to
$\bar N_{\!\wkR(\Lambdab_1+\rhob)-\rhob,\Lambdab_2...\Lambdab_\mn}$. Hence 
only finitely many $\wkr$ can contribute and we obtain a manifestly integral 
expression for $\dim\B$ in terms of multiplicities of invariant tensors of 
\findim\ \gb-modules. (In the special case of $m\eq3$ this description of the 
WZW fusion rules is known as the Kac\hy $\!$Walton formula \cite{KAc3,walt3}.)
As we will see in Subsection \ref{seul}, the \rhs\ of \erf{eN} can also be
interpreted as the Euler characteristic of a certain complex of \gb-\coin s.
 
If we so wish, we can proceed by expressing the number 
$\bar N_{\!\mub_1\mub_2...\mub_l}$ through weight multiplicities of \findim\ 
\gb-modules. To this end one has to implement the so-called Racah\hy Spei\-ser 
formula for the number of 
invariant tensors in a tensor product of \findim\ \gb-modules. We find
  \be  \bearll  \dim\B \!\! &= \dstyle\sumwwk\eps(\wkr)\sumwwb\epswb\,
  {\rm mult}_{\Lambdab_2...\Lambdab_{\mn-1}}(\wbar\wkr(\Lambdab_1+\rhob)\mi\rhob
  \mi\Lambdab_\mn) \\{}\\[-.8em]
  & \equiv \dstyle\sumww\epsw\, {\rm mult}_{\Lambdab_2...\Lambdab_{\mn-1}}
  (w(\Lambdab_1+\rhob)\mi\rhob\mi\Lambdab_\mn)  \,,  \eear \ee
where ${\rm mult}_{\Lambdab_2...\Lambdab_{\mn-1}}(\mub)$ denotes the 
multiplicity of the weight $\mub$ in the tensor product module
$\bar{\cal H}_{\Lambdab_1}\otimES\bar{\cal H}_{\Lambdab_2}{\otimes}\cdots\otimES
\bar{\cal H}_{\Lambdab_{\mn-1}}$.

\subsection{Resummation}

Alternatively, we can also express $\,\dim\B\,$ completely in terms of elements
of the modular matrix $S$ that describes the transformation properties 
of the characters of the \g-modules \hl\ under the modular transformation
$\tau\mapsto-1/\tau$. To this end, we recall that the affine Weyl group $W$
can be regarded as the semi-direct product of $\barW$ and of translations by 
elements of the coroot lattice $\lv$ of \gb. We can therefore write the 
elements of the affine Weyl group $W$ in \erf* as pairs $w=(\wbar;\betab)$ 
of elements $\wbar\iN\barW$ of the Weyl group of \gb\ and of coroot lattice 
elements $\betab\iN\lv$. Then the induced action \Erf wh of $w\iN W$ on the 
weight space of \gb\ is given by 
  \be  w(\bar\mu)=\wbar(\bar\mu)+\kv\betab \,.  \ee
Also taking care of the shift in the effective level that results from the 
shift of the weights by the Weyl vector $\rho$ (which has level \gv) in the 
Weyl group action, we can then rewrite \erf* in the form
  \be \bearll  \NNL \!\!
  &= \dstyle\Frac1{|\Wb|}\, \intT \chibarX\chibl\mn(\zhb)
  \\{}\\[-.8em]&\hsp{3.8}
  \cdot \uvcb^{-1}(-\zhb) \dstyle\sumwwb\epswb\, \eE^{\wbar(\Lambdab_1+\rhob)}
  (\zhb) \sum_{\betab\in\lV} \eE^{(\kV+\gV)\betab}(\zhb) \,. \eear \Labl dj

Relation \erf{dj} expresses the dimension of the space of 
\cb s as the integral of a certain function over a torus, namely the maximal
torus of the \Lie\ $\barG$. We will now show that one can replace this
integration by a finite summation over suitable \gb-weights.
This is achieved with the help of (a slightly generalized version of)
Poisson resummation which states that
for any function $f{:}\; (\gbodual)_\reals\cong\reals^{\rank\gB}\to\complex\,$
that is periodic \wrt \lv\ and for any natural number $l$, the formula
  \be  \intT\! \sum_{\betab\in\lV}f(\hb)\, \eE^{\betab}(\zpi l\hb)
  = |l\lv/\lw|\!\! \sum_{\lambdab\in l^{-1}\lW/\lV}\!\! f(\lambdab) \labl{lem}
holds. Here as before $\dt$ is the normalized Haar measure on $\barH$ and
$\hb\iN\gbo$ is related to $\gh\iN\barH$ by $\gh\eq\Exp(\zhb)$. Also, the 
integrand of the \lhs\ is single-valued on $\barH$ because $f$ is by assumption
\lv-periodic, and because for different elements 
$\hb_1,\,\hb_2\iN\gbodual$ 
which satisfy $\Exp(\zhb_1)\eq\gh\eq\Exp(\zhb_2)$
the values of $\betab(\hb_1)$ and $\betab(\hb_2)$ for $\betab\iN\lv$ 
differ by an integer, the exponential function is \lv-periodic as well.

To derive relation \erf{lem}, we introduce the Fourier transform $\tilde f$
of $f$. The Fourier components are labelled by the lattice dual to \lv, i.e.\
by the weight lattice \lw:\,%
 \futnote{We identify \gbo\ and the weight space \gbodual\
via the non-degenerate invariant bilinear form.}
$f(\hb)\eq\sum_{\lambdab\in\lW} \eE^{2\pi\ii\lambdab(\hb)} \tilf_\lambdab$.
Then the \lhs\ of \erf{lem} can be written as
  \be  \intT\! \sum_{\betab\in\lV}\sum_{\lambdab\in\lW}
  \tilf_\lambdab \, \eE^{2\pi\ii (\lambdab+l \betab)(\hb)}
  = \sum_{\lambdab\in\lW} \tilf_\lambdab \sum_{\betab\in\lV}
  \intT \eE^{\zpi(\lambdab+l\betab)(\hb)}
  = \sum_{\lambdab\in l\lV} \tilf_\lambdab \,.  \labl{ls}
Here the last equality holds because the integral over the torus is non-zero 
only when the integrand is constant, which happens only when $\lambdab=
-l\betab$, i.e.\ when $\lambdab$ lies in $l$ times the coroot lattice.
On the other hand, for the \rhs\ of \erf{lem} we compute
  \be  \sum_{\lambdab\in l^{-1}\lW/\lV}\!\! f(\lambdab) =
  \sum_{\mub\in\lW} \tilf_\mub \!\!\sum_{\lambdab\in l^{-1}\lW/\lV}\!\!
  \eE^{2\pi\ii(\lambdab,\mub)} = \mbox{\Large$|$} \Frac{\lW}
  {l^{\phantom|}_{\phantom|}\lVV} \mbox{\Large$|$}\!\sum_{\mub\in l\lV}
  \tilf_\mub\,, \labl{rs}
where we use the fact that the sum over $\lambdab$ only gives
contributions if the summand is constant and hence equal to one, which
happens precisely if the inner product $(\lambdab,\mub)$ is an integer.
The resummation formula \erf{lem} now follows by equating
relations \erf{rs} and \erf{ls}.

Applying now the result \erf{lem} to formula \erf{dj} for $\dim\B$, we get
  \begin{eqnarray}
  & \hsp{-.5} \dim\B \!\!&= \hllw \!\dsum_{\lambdab\in\hhE^{-1}\lW/\lV}\!\!\!\!
  \uvcb^{-1}(-\lambdab)\,
  \barchi_{\Lambdab_2}(\lambdab)\, \barchi_{\Lambdab_3}(\lambdab)\cdots
  \barchi_{\Lambdab_\mn}(\lambdab) \!\sum_{\wbar\in\barW} \!\eps(\wbar) \,
  \eE^{\zpi \wbar(\Lambdab_1+\rhob,\lambdab)}  \nonumber\\[.3em]
  &&= \hllw \!\dsum_{\lambdab\in\hhE^{-1}\lW/\lV}\!\! J(\lambdab)\,
  \barchi_{\Lambdab_1}(\lambdab)\, \barchi_{\Lambdab_2}(\lambdab)\cdots
  \barchi_{\Lambdab_\mm}(\lambdab)\, \barchi_{\Lambdab_\mn}(\lambdab)
  \,, \label{dp}\end{eqnarray}
where in the second line we used once more the Weyl character formula.

\subsection{Characters and the modular matrix $S$}

The final step in establishing the Verlinde formula
is now to express the result \erf{dp} in terms of the entries of the modular 
matrix $S$ that governs \cite{KAc3} the transformation of the affine characters
$\chii_\Lambda$ ($\Lambda\iN\pk$) under the modular transformation
$\tau\mapsto-1/\tau$. This is possible because the generalized quantum 
dimensions, i.e.\ ratios of $S$-matrix elements, coincide with the characters 
of the horizontal subalgebra, evaluated at suitable arguments.

We first rewrite the summation in expression \erf{dp}, which extends over 
the \gb-weights
$\lambdab\iN(\hhE^{-1}\lw)/\lv \equiv \lw/(\hhE\lv)$, in such a way that only
weights in the dominant Weyl alcove appear. The function $J$ \Erf Jh is the 
square of a function that is odd under the Weyl group $\barW$ and hence
vanishes when $\hb$ is on the boundary of some Weyl
chamber. Owing to the $\barW$-invariance of the characters and of $J$,
it follows that each Weyl chamber gives the same contribution to the sum. 
Restricting the summation to those weights that belong to the dominant Weyl
alcove therefore just amounts to cancelling the prefactor $1/|\barW|$. In the 
summation we are then left with the weights of the form
  \be  \lambdab = \frac{\Lambdab+\rhob}\hhe \,,  \labl4
where the weights $\Lambdab+\rhob$ are the integral weights in the interior of
the dominant Weyl alcove at level $\kv+\gv$. The weights $\Lambdab$ are thus
the integral weights in the dominant Weyl alcove at level \kv, i.e.\ precisely
the elements of the set $\pk$ \erf{pk}. Formula \erf{dp} can therefore
be rewritten as
  \be  \dim\B= \hllw\sumk\Lambdab J\Lrkg\,\barchi_{\Lambdab_1}\LrkG\,\barchi_
  {\Lambdab_2}\LrkG \cdots \barchi_{\Lambdab_\mn}\LrkG \,. \labl{dq}

A closed expression for the modular matrix $S$ is provided by the \kpf\
  \be  \bearl  S_{\Lambda,\Lambda'}^{}=(-\ii)^{(\dim\gB-\rank\gB)/2_{}}\,
  |\lW/\lV|_{}^{-1/2}(\kV+\gV_{})^{-{\rm rank}\,\gB/2} \\{}\\[-.66em] \hsp{9}
  \cdot \dsum_{\wbar \in\barW} \eps(\wbar)\,
  \exp\lLb -\Frac{2\pii}{\kV+\gV}\,(\wbar(\Lambdab+\rhob)
  ,\Lambdab'+\rhob) \lRb \,.  \eear  \labl{kpf}
By comparison with the Weyl character formula, it follows that the characters
$\barchi_{\Lambdab_i}$ evaluated at the weights \erf4 satisfy
  \be  \barchi_{\Lambdab_i}\Lrkg =
  \LlB \frac{S_{\Lambda_i,\Lambda}}{S_{\klo,\Lambda}} \LrB^* \,. \labl i
Similarly, together with the denominator identity one finds that
  \be \bearll
  S_{\klo,\Lambda} \!\!&= {|\lw/\hhE\lv|}^{-1/2}_{}\displaystyle\prod_
  {\alphab>0} 2 \sin \Frac{(\bar\Lambda+\bar\rho,\alphab)\pi}{\hhe}  \\[.8em]
  &= \ii^{-|\bar\Delta_+|}\, {|\hhE\lv/\lw|}^{1/2}_{} \displaystyle\prod
  _{\alphab>0} \Llb \!\exp\lLb \ii\pi\alrkg \lRb -\exp\lLb -\ii\pi\alrkg
  \lRb \Lrb  \,, \end{array}\ee
which in turn implies
  \be  |S_{\klo,\Lambda}|^2 =  \hll J\Lrkg  \,.  \labl0

We now insert the identities \erf i and \erf0 into formula \Erf dq for
$\dim\B$. We then finally obtain (using also the fact that $S_{\klo,\Lambda}$
and $\dim\B$ are real)
  \be \dim \B^{}_{\{\Lambda_i\},\{p_i\}} =
  \sumk\Lambdab (S_{\klo,\Lambda})^2 \, \prod_{i=1}^{m}
  \frac{S_{\Lambda_i,\Lambda}}{S_{\klo,\Lambda}} \,. \labl V
This is our desired result, the Verlinde formula that expresses the 
dimension $\dim\B$ of spaces of genus zero chiral WZW blocks through the
entries of the matrix $S$ \erf{kpf}. Note that the \rhs\ of \erf V is 
typically a complicated combination of complex numbers (actually all entries 
of $S$ lie in a cyclotomic extension of the rationals), i.e.\ unlike the 
alternative expression on the \rhs\ of \Erf eN, it is not manifestly integral.

\Sect{Homological aspects of the Verlinde formula}S

In this final section we add a few comments on general aspects of the Verlinde 
formula and present an interpretation of the Verlinde multiplicities $\dim\B$
as expressed by formula \erf{eN} in terms of Euler characteristics of certain 
complexes of \coin s.

\subsection{The Verlinde formula}

In the literature, derivations of the Verlinde formula have been described
at various levels of rigor, ranging from more heuristic considerations
to mathematically complete proofs. It seems fair to say that typically
the more rigorous these proofs are, the smaller the number of situations is to
which they apply. A \cft\ deduction which does not use any specific properties 
of the chiral symmetry algebra of the theory, and hence applies to arbitrary 
rational \cfts, was given in \cite{verl2,mose,bryz};
it involves certain formal manipulations with chiral blocks and hence may
be regarded as somewhat heuristic. 
All other proofs known to us involve (implicitly or explicitly) the block 
algebra $\gUU$,  and thereby the \rep\ theory of affine \lie s; \asac\ they
work exclusively for WZW \cfts. 
(The two kinds of approaches to the Verlinde formula also differ drastically in
another aspect: those applicable to arbitrary \cfts\ only aim at 
establishing a relation between the dimensions of spaces of chiral blocks and 
the modular transformation properties of the characters, i.e.\ the zero point 
blocks at 
genus one, but they do not predict any concrete expressions for these modular
transformations. In contrast, the proofs that only work in the WZW setting do
provide such expressions, in particular the explicit form of the modular
transformation matrix $S$; roughly speaking, they combine the information 
encoded in the Verlinde formula for general \cfts\ with the \kpf\ \erf{kpf} 
for $S$.) Among these approaches which apply to the WZW case, there is one 
\cite{dawe,blth} that is based on the path-integral 
quantization of \cS gauge theories and 
is accordingly affected by the usual difficulties in setting 
up a rigorous quantization procedure, e.g.\ concerning the proper mathematical
setting for path integrals. A related approach \cite{witt27,axdw} combines 
holomorphic quantization of \cS theories and surgery manipulations on 
three-manifolds. There is also an approach \cite{fink} which combines the 
isomorphism \cite{kalu3+4,kalu5+6} between the tensor categories of certain 
modules over affine \lie s \g\ at negative level and categories of modules 
over quantum groups $\U_q(\gM)$ with $q$ a root of unity with an isomorphism 
between categories of \g-modules at positive and negative levels to deduce the 
Verlinde formula from the \rep\ theory of quantum groups. 

Another possibility is to exploit the fact that the spaces of chiral blocks are
isomorphic to the spaces of holomorphic sections in line bundles
over certain projective varieties, namely over moduli spaces of semi-stable
principal bundles with structure group $\barG$, where $\barG$
is the real, compact, connected and simply connected Lie group with \lie\ \gb. 
(In the mathematics literature, sometimes this isomorphism between chiral 
blocks and geometric objects, rather than an expression for their common 
dimension, is 
meant when the term `Verlinde conjecture' is used.) One can attempt
to analyze these spaces with various methods of traditional \findim\ algebraic
geometry; this way several proofs have been established for the case of \wzwts\
based on $\gB=\sltwo$, where the vector bundles have rank two and their moduli
space can be described rather explicitly (see e.g.\ \cite{besz,thad2,jewe}, as
well as \cite{sorg}\,%
 \futnote{Other sources for references are \cite{beau2} and \cite{SCho2},
and also the WWW pages\\ {\tt http:/$\!$/www.ictp.trieste.it/\~{}mblau/ver.html}
\,and\, {\tt http:/$\!$/www.desy.de/\~{}jfuchs/Vfcb.html}\,.}
for a more detailed exposition and for more references). Finally, using 
the methods of algebraic geometry, proofs that work for $\sln$ with 
arbitrary $N$ have been obtained in \cite{falt,beLa,beau}; these arguments start
from the description of the moduli space of flat connections over a curve
in a double coset form. The proof of \cite{falt} which uses torsion-free 
sheaves even applies to all simple \lie s \gb, except for 
$\gB=F_4$, $E_6$, $E_7$ 
or $E_8$. While these proofs are formulated in a purely algebraic setting, a 
derivation where topological tools play an essential r\^ole was given
in \cite{tele}; there the modules are completed to Hilbert spaces 
and the Verlinde formula is obtained from certain vanishing theorems for 
complexes of these Hilbert spaces (compare subsection \ref{seul} below).

The argument that we presented in sections \ref{s.T}\,--\,\ref{s.F}
makes extensive use of the \rep\ theory of affine \lie s and is 
in most aspects rather different from all those mentioned above. In 
particular, all our arguments work simultaneously for \wzwts\ based on 
arbitrary simple \lie s \gb;\,%
 \futnote{This applies likewise to the vanishing theorem in \cite{tele}.}
at no step in the derivation is there any need to distinguish between 
different cases that have to be treated separately. Furthermore,
when combined with the results of \cite{fusS6} and \cite{fusS4}, \resp, our
methods should provide a possibility to characterize the \scbs\ and to prove the
Verlinde formula not only for \wzwts, but also for two other classes of
\cfts. The first of these classes is given by the so-called integer spin simple
current extensions of \wzwts; these correspond to such
\wzwts\ which in a Lagrangian setting are associated to
non-simply connected group manifolds, while for ordinary \wzwts\ one always
deals with the simply connected covering group of the simple \lie\ \gb.
The second class of theories consists of all coset \cfts, including in
particular those coset
theories in which so-called field identification fixed points are present.
As a final application we mention that the present representation theoretic
approach, combined with the results obtained in \cite{fusS3,furs},
should be helpful in verifying the conjectures made in \cite{fuSc8} concerning
the values of certain traces on the spaces of \cb s. 
(These traces have interesting applications to the construction of \cb s for 
integer spin simple current extensions of \wzwts, to the classification
of boundary conditions for \cfts\ \cite{fuSc5}, and in algebraic geometry 
to the derivation of a Verlinde formula for non-simply connected groups.)

\subsection{Complexes of \coin s}

We will now present a  homological interpretation of formula \erf{eN} which 
expresses the Verlinde
multiplicities $\dim\B$ as an alternating sum of non-negative integers. To this 
end we use the BGG resolution of an irreducible highest weight module of
\g\ in terms of parabolic Verma modules \erf{vuu} which are free 
$\g^-$-modules. We introduce the (finite) direct sums
  \be  \Verl^\jj:= \sumwlj \Verm^{}_{\!\wP(\Lambda+\rho)-\rho}  \ee
of generalized Verma modules.
Here $\ell(w)$ denotes the length of the Weyl group element $w$.
Note that the modules $\Verl^\jj$ are still free $\gminus$-modules, and that
for every integrable \g-weight $\Lambda$ and every $\wp\iN\Wp$ the \gb-weight
$w(\Lambdab\pl\rhob)\mi\rhob$ is dominant integral. In particular we have
  \be  \Verl^\jj \,\cong\, \U(\gminus) \otimes \llb\!\! \sumwlj
  \! \hilb_{\wP(\Lambdab+\rhob)-\rhob} \lrb
  = \U(\gminus) \otimes \hilb^{(j)}_\Lambda   \ee
as $\gminus$-modules, where $\hilb^{(j)}_\Lambda$ is the \findim\ \gb-module
  \be \hilb^{(j)}_\Lambda:= \sumwlj \hilb_{\wP(\Lambdab+\rhob)-\rhob} \, . \ee
 
The {\em parabolic BGG resolution\/} of an \ihwm\ \hl\ over \g\ with integrable
\hw\ $\Lambda$ states \cite{gale} that there exists a semi-infinite exact 
sequence of \g-modules with \g-module homomorphisms of the form
  \be  \ldots \to \Verl^\jj \to \Verl^\jM \to \ldots \to \Verl^\jo
  \equiv \Verl^{} \to \hl \to 0 \,. \labl{133}
Note that this complex is governed by the same set $\Wkr$ of representatives
of the coset $W/\barW$ that we encountered in Subsection \ref{dbf}.
When we tensor the sequence \erf{133} with the \findim\ $\gb$-module
$\barB\equiv\bigotimes_{i=1^{}}^{\mm_{}}\hlib$, we obtain a
semi-infinite exact sequence
  \be \bearl  \ldots \to \BarB\otimeS \Verm^\jj_\Lambdam \to
  \BarB\otimeS \Verm^{(j-1)}_\Lambdam \to \ldots \\[.7em] \hsp8 \ldots \to
  \BarB\otimeS \Verm^{(1)}_\Lambdam \to
  \BarB\otimeS \Verm_\Lambdam \to \BarB\otimeS \hil_{\Lambdam} \to 0
  \,. \eear  \labl{tgse}
We now observe that (by using the isomorphism \erf{47} to replace also the 
first affine module in the description \erf{biso} of the \cb s by a \findim\ 
\gb-module) the space \B\ of \cb s is isomorphic to the \coin s of 
$\BarB\otimeS \hil_{\Lambdam}$ \wrtt \bla\ $\tildeg$ for a single insertion 
point. ($\tildeg$ is the algebra of \gb-valued algebraic functions that are 
allowed to have a pole of finite order at a single distinguished point
and is hence isomorphic to $\gb\oplus\g^-$.)
We would like to combine this information with the exact sequence \erf{tgse}.
To this end, we first observe that the spaces in the sequence \erf{tgse}
(except for the last two) carry the structure of $\tildeg$-modules which 
are free $\g^-$-modules.
On the other hand, taking \coin s of an exact sequence does in general not 
produce another exact sequence, but only a complex. However, one can show (see 
Appendix \ref{arie}) that this procedure still constitutes a right-exact 
functor, so that the exact sequence \erf{tgse} provides us, by taking \coin s
\wrt $\tildeg$, with a semi-infinite complex
  \be \bearl  \hsp{-1.1}
  \ldots \to \coi{\BarB\otimeS \Verm^\jj_\Lambdam}\tildeg \to
  \coi{\BarB\otimeS \Verm^{(j-1)}_\Lambdam}\tildeg \to \ldots \\[.7em]
  \hsp{5.5} \ldots\to \coi{\BarB\otimeS \Verm^{(1)}_\Lambdam}\tildeg \to
  \coi{\BarB\otimeS \Verm_\Lambdam}\tildeg \to
  \coi{\BarB\otimeS \hil_\Lambdam}\tildeg  \equiv \B \to 0    \eear \labl{res0}
of spaces of $\tildeg$-\coin s that is still exact at the two right-most 
positions.

Next we note that using the two fundamental facts \erf{hhh} and \erf{17} about 
\coin s of free modules, one can derive the isomorphism
  \be \bearll \coi{\BarB\otimeS \Verm_{\!\mu}}\tildeg \!\!&
  \;\cong\;  \coi{\BarB\otimes (\ugmi\otimeS\hilb_\mub)}\tildeg \\{}\\[-.7em]&
  \;\cong\;  \coi{\ugmi \otimes (\BarB\otimeS \hilb_\mub)}{\gminus\oplus\gB}
  \;\cong\;  \coi{ \BarB\otimeS \hilb_\mub}\gB  \eear \ee
of vector spaces. Hereby we can simplify the various spaces that occur in the 
complex \erf{res0}, so that it can be written as a complex that, except for 
$B\eq\coi{\BarB\otimeS \hil_\Lambdam}\tildeg$, involves only \gb-\coin s:
  \be \hsp{-5.6} \bearl  \ldots \to \coi{\BarB\otimeS \hilb^\jj_\Lambdam}\gB \to
  \coi{\BarB\otimeS \hilb^{(j-1)}_\Lambdam}\gB \to \ldots \\[.7em]
  \hsp{7.8} \ldots\to \coi{\BarB\otimeS \hilb^{(1)}_\Lambdam}\gB \to
  \coi{\BarB\otimeS \hilb_\Lambdam}\gB \to \B \to 0  \,.   \eear \labl{erf}

Thus we have finally arrived at a complex of \gb-\coin s which is governed by 
the subset $\Wp$ of the Weyl group $W$ of \g. Moreover, by the
same arguments as in the previous section, we deduce from formula \Erf0f 
for \gb-\coin s that this semi-infinite complex is not really infinite, but 
actually constitutes a finite complex of \gb-\coin s, i.e.\ we even have
  \be \bearl  0 \to \coi{\BarB\otimeS\calhb^{(\jm)}_{\Lambdab_m}}\gB \to
  \coi{\BarB\otimeS\calhb^{(\jm-1)}_{\Lambdab_m}}\gB \to \ldots  \\[.7em] \hsp9
  \ldots \to \coi{\BarB\otimeS\calhb^{(1)}_{\Lambdab_m}}\gB \to
  \coi{\BarB\otimeS\hilb_{\Lambdab_m}}\gB \to \B \to 0   \eear \labl{fin}
for some non-negative integer $\jm$. 

\subsection{Verlinde multiplicities and Euler characteristics}\label{seul}

Owing to the finiteness of the complex \erf{fin}, we can express
the dimension of the space \B\ of chiral blocks as
  \be  \dim\B = \eul + \sum_{j=0}^\jm (-1)^j \, \dim \llb \coi{\BarB{\otimes}
  \calhb^{(j)}_{\Lambdab_m}}\gB \lrb  \ee
through the Euler characteristic \eul\ of the complex \erf{fin} and an alternating
sum of dimensions of spaces of \gb-\coin s. The latter coincide with the
number of singlets in the respective \gb-modules. Together with the identity 
$(-1)^{\ell(w)}\eq\epsw$ it then follows that the result \Erf eN for the
Verlinde multiplicities can be rephrased as the statement that the Euler
characteristic \eul\ of the complex \erf{fin} vanishes.

Actually, we conjecture that not only $\eul=0$, but that 
the whole homology of the complex is zero, i.e.\ that
\erf{fin} is in fact an {\em exact\/} sequence. Note that
it is immediate that the sequence 
is indeed exact as long as the number of insertion points is $m\leq2$.
In these cases the only non-vanishing spaces in the complex \erf{erf} are those
involving the modules $\VV_\Lambdam$ or $\hil_\Lambdam$ (for $m\eq2$ this is
easily seen by considerations similar to those that led to formula \Erf2d;
the case $m\eq1$ can be regarded as a special case of the $m\eq2$ situation
where the second insertion point carries the weight $\Lambda\eq\klo$).
Exactness then already follows from the fact (compare Appendix \ref{arie}) that 
taking \coin s constitutes a right-exact functor.
Another situation where exactness can be established directly (though in a
somewhat lengthy manner, by using the explicit form of the BGG maps and of the
Clebsch\hy Gordan decomposition) is $m\eq3$ for $\gB\eq A_1$.

This conjecture about the vanishing of the homology of \erf{fin} has also
been made, from a different perspective, in \cite{kuma5}.
We do not know of any direct Lie algebraic proof of our conjecture. However, a 
similar result has been obtained in \cite{tele}, based on a completion of the 
highest weight modules \hl\ to topological Hilbert spaces. The vanishing 
theorem of \cite{tele} refers to the cohomology of a complex that involves
dual space of the space $\coi\hvl\gcPM$ of \coin s considered here, i.e.\
the space of $\gcPM$-invariants in the algebraic dual $(\hvl)\dual_{}$
of $\hvl$. The vanishing statement for the homology should be related, of 
course, to the corresponding statement for the cohomology in \cite{tele}.
Conversely, a proof of our conjectures concerning the structure of the completed
tensor product module \hhc\ in subsection \ref{sB}, and hence of the
vanishing of the Euler number of the complex \erf{fin}, might constitute a first
step towards a more direct derivation of the vanishing theorem of \cite{tele}.
In this context it could also be interesting to analyze 
the lowest weight socle of the algebraic dual \hhd\ of \hh\
(compare appendix \ref{sscl}).

We regard the results of \cite{tele} also as an indication that unitarity 
should be a crucial input for the vanishing of the whole homology. 
Indeed, this fits nicely with the observation \cite{mawa2}
that when one inserts the modular matrix $S$ for so-called admissible 
\cite{kawa2} \g-modules (which appear in \wzwts\ at fractional level and are 
not unitarizable), one still obtains integers, which however
can now also be negative,\,%
 \futnote{The proper definition of $S$ in these cases is, however, problematic;
compare e.g.\ section 5 of \cite{dolm4}.}
and as already noted in \cite{fema3},
this suggests that in this situation the Verlinde multiplicities may still 
possess a homological interpretation.
More specifically, note that when one deletes the space $B$ in the complex 
\erf{fin} one obtains another complex whose Euler characteristic does not
vanish any more but is given by $\dim\B$; in the non-unitary case we would
expect the presence of a similar complex whose Euler characteristic is still
given by the Verlinde multiplicity, but whose
homology is no longer concentrated at the last position.

The interpretation of such a complex and its relation to the true fusion 
coefficients (which are dimensions and therefore are manifestly non-negative,
and which can for example be deduced by explicitly
solving the constraints implied by null vector decoupling, in particular the
Knizh\-nik\hy Za\-mo\-lod\-chi\-kov equation \cite{fema3,awya,pery3,fugp6})
remains to be clarified. In this context the observation \cite{mawa2} that 
in many cases the Verlinde
multiplicities are controlled by the integral parts of the weights, and hence
by fusion rules of unitary theories, is particularly interesting. 
In accordance with general experience with \bwb theory, it might 
find its explanation by the property that in the non-unitary case the  
homology is still concentrated
at a single position of the complex, albeit not at the last one.

We finally remark that the traces of the action of certain outer automorphisms
on the spaces of \cb s \cite{fuSc8} are (possibly negative) integers.
This suggests that they might find a natural interpretation as (possibly
twisted) Euler characteristics as well. Such an interpretation would not
only explain the rather surprising integrality properties of these traces,
but may also be helpful for a proof of the conjectures in \cite{fuSc8}.

\newpage
\appendix

\sect{Further properties of \coin s}\label{acoin}

\subsection{\Coin s of free modules}

When $W$ is a {\em free\/} module over some \lie\ $\h$, i.e.\ when
  \be  W\,\cong\,\U(\h)\otimes X \,\equiv\,\U(\h)\otimesc X   \labl{UX}
for some $\complex$-vector space $X$, then the tensor product $V \otimeS W$
of $W$ with any other $\h$-module $V$ is again a free $\h$-module.
More precisely \cite[Prop.\,1.7]{gale}, there exists a natural isomorphism
  \be  V \otimes (\U(\h)\otimeS X) \;\cong\; \U(\h) \otimes (V\otimeS X)
  \labl{17}
of $\h$-modules (on the \rhs, $V\otimeS X$ is a tensor product of vector
spaces); the isomorphism $\U(\h)\otimeS(V{\otimes}X)\,{\congto}\,V\otimeS
(\U(\h){\otimes}X)$ is given by
  $    u\oT(v\otim x) \mapsto \sum_\ell u^1_\ell v\oT(u^2_\ell\otim x)$,
where $\sum_\ell u^1_\ell\otim u^2_\ell=\Delta(u)$ is the coproduct of
$u\iN\U(\h)$. 

Moreover, by taking \coin s the result \erf{17} implies that
there is a natural isomorphism
  \be  \coi{V \otimeS W}\h \;\cong\; V\otimeS X  \labl{VX}
of vector spaces. In other words, the elements of the vector space $V\otimeS X$ 
provide natural representatives for the \coin s of $V\otimeS W$.

Next we consider the situation that the \lie\ $\h$ is the
semi-direct sum $\h=\h_1\,{\oplus}\,\h_2$ of an ideal $\h_1$ and a subalgebra
$\h_2$. Then $\coi V{\h_1}$ is a $\U(\h_2)$-module, and one can evaluate
the space $\coi V\h$ of \coin s in a two-step procedure:
  \be  \coi V\h={\lfloor\coi V{\h_1}\rfloor}_{\h_2}^{} \,.  \labl{hhh}
Furthermore, when $V$ is a free $\h$-module, then equality \erf{hhh} already 
holds when $\h=\h_1\oplus\h_2$ with $\h_1$ and $\h_2$ subalgebras, i.e.\ it 
is not required that $\h_1$ is an ideal. 

In order to establish these statements, we first
note that whenever $\h\eq\h_1\,{\oplus}\,\h_2$ as a direct sum of vector
spaces, then it follows from the Poin\-ca\-r\'e\hy Birk\-hoff\hy $\!$Witt 
theorem (upon choosing a basis of $\h$ that is the union of bases 
$\{h_{\sss(1)}^a\}$ of $\h_1$ and $\{h_{\sss(2)}^q\}$
of $\h_2$ to obtain a suitably ordered basis of $\U(\h)$)
that every $u\iN\U(\h)$ can be written as
   \be  u = \xi\,\bfe + \sum_a\xi_a h_{\sss(1)}^a + \sum_q\eta_q h_{\sss(2)}^q
   + \sum_{a,q}\zeta_{a,q} h_{\sss(1)}^a h_{\sss(2)}^q + \ldots \,,  \labl{u12}
and that this decomposition is unique. In particular, $\Up(\h)$ decomposes as
a vector space as
  \be  \Up(\h) = \Up(\h_1) \oplus \U(\h_1)\,\Up(\h_2) \,. \labl{UUU}
Now in the special case when $\h$ is the
semi-direct sum $\h\eq\h_1\,{\oplus}\,\h_2$ of an ideal $\h_1$ and a subalgebra
$\h_2$, it is a direct consequence of $[\h_1,\h_2]\subseteq\h_1$
that $\coi V{\h_1}$ is a $\U(\h_2)$-module. Together with \erf{UUU} it
then follows that for any $\h$-module $V$ one can evaluate the space
$\coi V\h$ of \coin s in a two-step procedure as described in \erf{hhh}.

Furthermore, when $V\cong\U(\h)\otimeS X$ is a free $\h$-module, then
the decomposition \erf{u12} of elements of $\U(\h)$ implies that
every element of $V$ can be written {\em uniquely\/} as
  \be  v = \bfe\oT x + \sum_i \uep i \oT x_i + \sum_j \uzp j \oT y_j
  + \sum_{i,j} \ueP i\uzP j \oT x_{ij}  \labl{vx}
(all sums finite) with $x,x_i,y_j,x_{ij}\iN X$, $\uep i,\ueP i\iN\h_1$
and $\uzp i,\uzP i\iN\h_2$. It follows that a natural representative
of the class of $v$ in $\coI V\h$ is given by $x\iN X$.
Moreover, when $\h_1$ is actually a subalgebra of $\h$, one can then consider
the space of \coin s $\coI V{\h_1}$, and a natural representative of the class
of $v$ in $\coI V{\h_1}$ is given by $\bfe\otim x+\sum_j\uzp j\otim y_j$. 
Finally, when also $\h_2$ is a subalgebra of $\h$ and $\coI V{\h_1}$ is an 
$\h_2$-module,
then we can take $\h_2$-\coin s of $\coI V{\h_1}$, and a natural representative
of the class of $\bfe\otim x+\sum_j\uzp j\otim y_j$ is again $x$.
Note that $\coI V{\h_1}$ can indeed be endowed with the structure of an
$\h_2$-module,
namely by identifying the action of $\h_2$ on classes by its action on the
distinguished representatives $\bfe\otim x+\sum_j\uzp j\otim y_j$, i.e.\
by demanding that $h_2\iN\h_2$ acts on the $\Up(\h_1)$-class
$[v]\iN\coI V{\h_1}$ of $v\iN V$ by
  \be  h_2\,[v] := [ h_2\otim x+\sum_jh_2\uzp j \otim y_j ]  \labl Y
when $v$ is decomposed as in \erf{vx}.
Using the uniqueness of that decomposition, it is straightforward to check
that the prescription \erf Y yields a linear \rep\ of $\U(\h_2)$ on
$\coI V{\h_1}$.

In summary, we have shown that indeed, when $V$ is a free $\h$-module and
$\h\eq\h_1\,{\oplus}\,\h_2$ as a direct sum of vector spaces,
then for the equality \erf{hhh} of spaces of \coin s to hold it is
sufficient that $\h_1$ and $\h_2$ are subalgebras of $\h$,
while for arbitrary $\h$-modules $V$ one must in addition require that
$\h_1$ be an ideal of $\h$.

\subsection{Right-exactness}\label{arie}

Here we demonstrate that the functor of taking \coin s is right-exact.
We first observe that whenever
  \be  \ldots \to V^{p+1} \fto{p+1} V^p \fto p V^{p-1} \to\ldots \to V^1
  \fto1 V^0 \fto0 0  \labl{orse}
is a semi-infinite exact sequence of modules of an arbitrary \lie\ $\h$, then
by taking \coin s one obtains an analogous complex
  \be  \ldots \to \coi{V^{p+1}}\h \fhto{p+1} \coi{V^p}\h \fhto p \coi{V^{p-1}}\h
  \to\ldots \to \coi{V^1}\h \fhto1 \coi{V^0}\h \fhto0 0   \labl{cose}
of vector spaces.
To define this complex we note that the maps $f^p{:}\; V^p \to V^{p-1}$ are
$\h$-in\-ter\-twi\-ners. Therefore they give rise to linear maps
$f_\h^p{:}\;\coi{V^p}\h \to \coi{V^{p-1}}\h$ between the spaces of \coin s with
respect to $\h$, which satisfy $f^p_\h\Circ\pi^p=\pi^{p-1}\Circ f^p_{}$,
where for each $p$ the map $\pi^p{:}\; V^p\to\coi{V^p}\h$ is the canonical
projection; thus they act as $f_\h^p{:}\; [v^p]\mapsto[f^p(v^p)]$,
where for $v\iN V$ we denote by $[v]$ the equivalence class of $v$ modulo 
$\Up(\h)V$, and this action does not depend on the choice of representatives.

Let us verify that \erf{cose} is indeed a complex. If
$[v^p] \iN \Im f^{p+1}_\h$, then there exist a $w^{p+1}\iN V^{p+1}$ as well as
finitely many elements $w_j^p\iN V^p$ and $x_j\iN\h$ such that $v^p$ can be
written as $v^p=f^{p+1}(w^{p+1})+ \sum_j x_j w_j^p$.
The facts that the sequence \erf{orse} is exact and that $f^p$ intertwines
the $\h$-action therefore imply that $f^p(v^p)=f^p(\sum_jx_jw^p_j)=
\sum_jx_jf^p(w^p_j)$. It follows that $[f^p(v^p)]=0$,
so that $[v^p]\iN\Ker f^p_\h$. Hence we have
  \be  \Im f^{p+1}_\h \subseteq \Ker f^p_\h \,, \Labl12
or in other words, $f^{p}_\h \Circ f^{p+1}_\h = 0$, for all $p\iN\zetplus$.

We can now show that the complex \erf{cose} is always {\em right-exact\/}, i.e.\
exact at its last two entries. (In general, there is however no reason why 
this complex should be exact also at the other positions.)
We start at the right of the diagram. The map $f^1$ is surjective,
so that we have $\Im(f^1_\abar)=\coi{V^0}\h$. Since $\Ker f^0_\h=\coi{V^0}\h$,
this already shows exactness at $\coi{V^0}\h$. {}From the result \Erf12 for
general $p$ we also already know that $\Im f^2_\h \,{\subseteq}\, \Ker f^1_\h$.
To show the converse, suppose that $[v^1]\iN \Ker f^1_\h$. Then there exist
a finite number of $x_j\iN\h$ and
$w^0_j\iN V^0$ such that $f^1(v^1) = \sum_jx_j w^0_j$. Moreover, since $f^1$ is
surjective, for each $j$ we can write $w^0_j = f^1(w^1_j)$ for some 
$w^1_j\iN V^1$. It follows that $f^1(v^1\,{-}\sum_jx_j w^1_j)= 0$, and hence 
$v^1\,{-}\sum_jx_jw^1_j\iN\Ker f^1$. Since the original sequence \erf{orse} is 
exact at $V^1$, there then exists some $w^2\iN V^2$ such that $f^2(w^2) =
v^1\,{-}\sum_jx_j w^1_j$. Therefore $f^2_\h([w^2])=[v^1\,{-}\sum_jx_j w^1_j]= 
[v^1]$, and hence $[v^1]\iN\Im f^2_\abar$. Thus we have shown that 
$\Im f^2_\abar= \Ker f^1_\abar$, so that the sequence
\erf{cose} of \coin s is also exact at $\coi{V^1}\h$, as claimed.

\sect{Proof of the isomorphisms \erf{ihV} and \erf{hV}}\label{aproof}

Here we establish the isomorphisms \erf{ihV} and \erf{hV}. Important tools
are provided by the results about \coin s that were described in Appendix
\ref{acoin}.

We use the notations of subsection \ref{sBH}, and we pick a `global' coordinate 
$z$ of \pe\ such that the set $Q$ of insertion points contains the 
point $z\eq0$.  To derive \erf{ihV}, we first recall the isomorphism
$\Verl\cong\U(\gminus)\otimeS\hlb$ \erf{vuu}. Owing to the
general result \erf{VX}, upon tensoring with the $\gminus$-module $\neh$
\Erf ht this implies the natural isomorphism
  \be  \neh \otimes\hilb_\Lambdab \;\cong\;
  \coi{\neh \otimes \VV_{\Lambda}}\gminus \,.  \labl{chV}
Note in particular that according to \erf{chV} the space $\coi{\neh \otimeS
\VV_{\Lambda}}\gminus$ of \coin s is a $\gB(\pE{\setminus}Q)$-module. Next we 
observe that we can write the \lie\ $\gB(\pe{\setminus}(Q{\cup}\{\infty\}))$
as a vector space direct sum of the \lie\ $\gB(\pe{\setminus}Q)$ and the \lie\
  \be  \nealg:= \gbar\otimes z\,\complex\twobrac z  \,,  \ee
i.e.\ that there is an isomorphism
  \be  \gB(\pE\Setminus(Q\cup\{\infty\})) \,\cong\, \gB(\pE\Setminus Q)\oplus
  \nealg  \ee
of vector spaces, and moreover, that the two summands are in fact subalgebras of
$\gB(\pe{\setminus}(Q{\cup}\{\infty\}))$. Furthermore, a local coordinate around
the additional insertion point $\infty$ is given by $t\eq 1/z$. Thus via
$\nealg$ we have included functions with poles at this insertion point, 
or in other words, $\nealg$ acts on the module \hl\ by elements of $\gminus$
(this is of course the reason why we chose the superscript `$-$' in
the notation $\nealg$). Note that every vector of \hl\ is annihilated by all 
but a finite number of generators of $\nealg$,
so that we actually have to deal only with finite series; 
accordingly we will identify $\nealg$ with $\gminus$ from now on.

By taking \coin s \wrt $\gB(\pE{\setminus}Q)$ in \erf{chV},
we can then conclude that \erf{ihV} is valid. When doing so,
we just have to apply the identity \erf{hhh}; recall that for the validity
of \erf{hhh} it is sufficient that $\gminus$ and $\gB(\pE{\setminus}Q)$ are
subalgebras, since $\Verl$ is a free $\gminus$-module.
\smallskip

In order to show \erf{hV} as well, we first introduce an action of 
$\gminus$ in an obvious manner, i.e.\ analogous to the action \erf{rxf} of the 
\bla; thus on a factor
$\hil_{\Lambda'_j}$ of $\nehz$ the element $\xb\otim f\iN\gminus$ acts as
$\xb\otim\fqj(\z_j)$ with $\fqj(\z_j)$ the Laurent series of $f$ at $q_j$, while
on a factor $\hilb_{\Lambdab_i}$ of $\nehe$ it acts via evaluation, i.e.\
by $\xb\otim \tilde\fpi(0)$ with $\tilde\fpi(\z_i)$ the power series expansion
of $f$ at $p_i$. Note that this way the space $\neh$ is {\em not\/} turned 
into a $\gB(\pe{\setminus}(Q{\cup}\{\infty\}))\,$-module.
Indeed, the residue theorem that in the case of \erf{rxf} forced
the central terms to cancel now no longer implies such a cancellation,
because the functions can have additional poles at $\infty$, so that
in the Lie bracket of $\xb\otim f$ and $\yb\otim g$ ($\xb,\yb\iN\gb$,
$f,g\iN{\cal F}(\pe{\setminus}(Q{\cup}\{\infty\})$)
we are left over with a central term proportional to
$-\kv\,\kappa(\xb,\yb) \summI \resinf(f{\rm d}g)$. 
(In physicists' terminology, this is formulated as follows. Every contour
encircling all insertions points except for the one at infinity can be
deformed to a contour around infinity. This contour, however, does not have 
the standard orientation, thus accounting for the minus sign of the level.)
Thus the \rep\ is only a projective one, and precisely as in the two-block 
case discussed in \sref C we have to work with a central extension 
$\hat\gB(\pe{\setminus}(Q{\cup}\{\infty\})$ of the \bla\
$\gB(\pe{\setminus}(Q{\cup}\{\infty\})$. 

Let us now denote by $\kerl$ the kernel of the canonical surjection 
$\pi{:}\ \Verl\to\hl$. The space $\kerl$ is generated by a single primitive 
null vector $w$; in terms of the \hwv\ $w_\Lambda$ of $\Verl$ it is given by
  \be  w=(\ezm)^{\kV-(\bar\Lambda,\ttV)+1}w_{\Lambda} \,,  \ee
where by $E^{\pm\ttA}\iN\gB$ we
denote the step operators in \gb\ that correspond to the highest \gb-root
$\ttA$ and to its negative, \resp. Now the fact that
the element $\ezp$ of $\gplus$ acts locally nilpotently 
on the space $\neh$ \Erf ht means that for each $v\iN\neh$ there exists
a positive integer $M\eq M(v)$ such that $(\ezp)^M_{}v\eq0$.\,%
 \futnote{Note that the expression $(E^{-\bar\theta}\otim t)^M v
\equiv (E^{-\bar\theta}\otim z^{-1})^M v$ is an element of 
$\gB(\pe{\setminus}(Q{\cup}\{\infty\})$ only if
$z^{-1}\iN{\cal F}(\pe{\setminus}(Q{\cup}\{\infty\}))$. This is indeed
satisfied because we have $0\iN Q$. It is here that our 
assumption about $Q$ being non-empty enters.}
Further, by commuting the element
$(\ezp)^M$ of $\U(\g)$ through $(\ezm)^{M+\kV-(\bar\Lambda,\ttV)+1}$ and
using the fact that the \hwv\ $w_\Lambda$ of $\Verl$ is annihilated by
$\gplus$, it follows that the vector $w':=(\ezp)^M(\ezm)^{M+\kV-(\bar\Lambda,
\ttV)+1}w_{\Lambda}$ is a non-zero multiple of $w$. This means that
there is a non-zero vector $\tilde w\iN \VV_{\Lambda}$
such that $w$ can be written as $w=(\ezp)^M_{} \tilde w$. Just as on $\neh$,
the algebra $\gB(\pe{\setminus}(Q{\cup}\{\infty\})$ only acts
projectively on $\kerl$, but the central element of the extension 
$\hat\gB(\pe{\setminus}(Q{\cup}\{\infty\})$ now acts with value \kv\
rather than $-\kv$.

It follows that for each vector $v\iN\neh$ we have, as an element of
$\neh\otimes^{}_{\U(\gB(\pE\setminus(Q\cup\{\infty\})))}\VV_{\Lambda}
\equiv$ $\coi{\neh\otimeS\VV_{\Lambda}}{\gB(\pE\setminus(Q\cup\{\infty\}))}$,
the identity
  \be  v\oT w = v\oT (\ezp)^M_{}\tilde w = (-1)^M\, \llb(\ezp)^M_{}v\lrb
  \oT \tilde w =0  \,. \labl{csr}
Since the vector $w$ generates the kernel $\kerl$ as a (projective)
$\gB(\pe{\setminus}(Q{\cup}\{\infty\}))\,$-module, this result holds analogously 
for every other element of $\neh\otimes\kerl$ as well. It follows that the 
image of $\neh\otimes\kerl$ under the surjection
$\id\times\pi{:}\ \neh\otimes\VV_{\Lambda}\to\neh\otimes\hil_{\Lambda}$ is
zero in $\coi{\neh\otimes\hil_{\Lambda}}{\gB(\pE\setminus(Q\cup\{\infty\}))}$.
In short, the kernel $\kerl$ does not contain any \coin s:
  \be  \coi{\neh\otimeS\kerl}{\gB(\pE\setminus(Q\cup\{\infty\}))} = 0\,.
  \labl{input}

Now we have a short exact sequence $0\to\kerl\stackrel\imath\to\Verl\stackrel
\pi\to\hl\to0$, which when
tensored with the space $\neh$ yields another exact sequence
  \be  0 \to \neh\otimeS\kerl \to \neh\otimeS\Verl \to \neh\otimeS\hl \to 0
  \,.  \labl{nehseq}
Upon taking \coin s \wrt the \lie\ $\gB(\pE{\setminus}(Q{\cup}\{\infty\}))$, 
this provides us with a complex
  \be  0 = \coi{\neh\otimeS\kerl}{\gB(\pE\setminus(Q\cup\{\infty\}))}
  \to \coi{\neh\otimeS\Verl}{\gB(\pE\setminus(Q\cup\{\infty\}))}
  \to \coi{\neh\otimeS\hl}{\gB(\pE\setminus(Q\cup\{\infty\}))} \to 0 \,. \Labl0e

The right-exactness of the functor of taking \coin s that was described
in subsection \ref{arie} now tells us that \Erf0e is in fact an exact sequence.
{}From \erf{input} we can therefore
conclude that the isomorphism \Erf hV is valid, as claimed.

\sect{On the lowest weight socle of \hhD}\label{sscl}

The {\em socle\/} of a module is by definition the linear hull of all its
irreducible submodules; it is in fact a direct sum of irreducible 
modules. Here we are concerned with the {\em highest\/}, \resp\
{\em lowest weight socle\/} $\soc\hil$ of 
a module $\hil$ over a \lie\ \H, defined as the linear hull of all its
irreducible highest (\resp\ lowest) weight submodules, which is a direct 
sum of irreducible highest (\resp\ lowest) weight modules. The situation of our 
interest is the one where \H\ is an affine \lie\ and $\hil=\hhd$, with
\hh\ the tensor product of the level-\kv\ \inthwm\ $\HL1$ and the level-0
evaluation module $\barB$, which was considered in subsection \ref{sB}
(where \H\ was realized as the centrally extended block \alg\ $\gtc$).

Recall that we describe the space $B$ of \cb s as the space $\coi V\h$ of 
\coin s of the tensor product 
  \be  V = (\hh)\otimeS \hilt_{\Lambda_m}  \Labl0V
of \hh\ with an irreducible lowest weight \H-module $\hlm$ at level $-\kv$.
Its dual space $B\dual$ then coincides with the space $(V\dual)^\h$ of 
\H-invariants (singlets) in the \alg ic dual $V\dual$ of $V$, i.e.\ the 
elements of $(V\dual)^\h$ are in one-to-one correspondence to functions 
on the \coin s. This holds 
because the kernel of every $\psi\iN(V\dual)^\h$ contains the submodule
$\Up(\h)(V)$ of $V$, so that by setting $\hat\psi([v]) := \psi(v)$
for each $\psi\iN(V\dual)^\h$ one defines a linear function on the space
$\coi V\h$ of \coin s, and vice versa. 

One can even show that the module \hh\ has trivial \hlw\ socles. In contrast, 
every element in the dual $\coidual V\h$ gives rise to a submodule of 
$\hhd$ that is isomorphic to the irreducible lowest weight module
$\hlm$. Indeed, given any $\hat\psi\iN\coidual V\h$, we can define
for every $v\iN\hlm$ the linear function $\psi_v\iN\hhd$ by
  \be  \psi_v(w) := \hat\psi([w\otimes v]) \ee
for all $w\iN\hh$. One can check that the action of \g\ on $\hlm$ 
precisely reproduces the action of \g\ on the subspace
  \be  \hhdpsi:= \{ \psi_v\iN \hhd \Mid v\iN\hlm \}  \ee
of \hhd, and vice versa. In other words,
the subspace \hhdpsi\ is isomorphic to $\hlm$
as an \H-module. Moreover, by construction this correspondence between
$\hat\psi$ and \hhdpsi\ is one-to-one, so \hhd\ contains the direct sum
  \be  \bigoplus_{\hat\psi\in\coidual V\h} \hhdpsi
  \,\cong\, \coidual V\h \otimes \hlm  \labl{ds}
as a submodule. Note that this is indeed a {\em direct\/} sum (because 
the modules \hhdpsi\ are minimal non-zero submodules of \hhd), which
shows in particular that
  \be  \mult{\socm\hhd}{\hlm} \,=\, {\rm dim}\,\coi V\h \,.  \labl=

Clearly, the relation \erf= implies that
  \be  \mult{\hhd}{\hlm} \,\ge\,{\rm dim}\,\coi V\h \,.  \labl:
This inequality gets strengthened to strict equality iff 
the lowest weight socle series of \hhd\ terminates after its first term.
By showing that such a termination occurs under appropriate conditions on the
evaluation module $\barB$ (so as to implement the fact that it originates from
integrable \H-modules), one would obtain an alternative
possibility to determine the branching coefficient $\NNL$.

\vskip3.5em{\small
\noindent{\bf Acknowledgements}\\ We thank W.\ Eholzer, E.\ Frenkel,
O.\ Gabber, I.\ Kausz, W.\ Soergel, C.\ Sorger and C.\ Teleman 
for helpful comments. We also acknowledge inspiring discussions with
the participants of the Oberwolfach Workshop `Verlinde formula and 
conformal blocks' (see {\tt http:/$\!$/www.desy.de/\~{}jfuchs/Vfcb.html}),
which was held within the RiP program that is supported by the 
Volkswagen-Stiftung.}

\vskip3em
\ifnum\draftcontrol=0 \newpage \fi
 
 \def\wb{\,\linebreak[0]} \def\wB {$\,$\wb}
 \def\Bi{\bibitem }
 \newcommand\Erra[3]  {\,[{\em ibid.}\ {#1} ({#2}) {#3}, {\em Erratum}]}
 \newcommand\BOOK[4]  {{\em #1\/} ({#2}, {#3} {#4})}
 \newcommand\J[5]   {\ {\sl #5}, {#1} {#2} ({#3}) {#4} }
 \newcommand\Prep[2]  {{\sl #2}, preprint {#1}}
 \newcommand\inBO[7]  {\ {\sl #7},
                      in:\ {\em #1}, {#2}\ ({#3}, {#4} {#5}), p.\ {#6}}
 \def\jf    {J.\ Fuchs}
 \def\adma  {Adv.\wb Math.}
 \def\anop  {Ann.\wb Phys.}
 \def\aspm  {Adv.\wb Stu\-dies\wB in\wB Pure\wB Math.}
 \def\comp  {Com\-mun.\wb Math.\wb Phys.}
 \def\foph  {Fortschr.\wb Phys.}
 \def\fuaa  {Funct.\wb Anal.\wb Appl.}
 \def\hepa  {Helv.\wb Phys.\wB Acta}
 \def\ihes  {Publ.\wb Math.\wB I.H.E.S.}
 \newcommand\geap[2] {\inBO{Physics and Geometry} {J.E.\ Andersen, H.\
            Pedersen, and A.\ Swann, eds.} \MD\NY{1997} {{#1}}{{#2}} }
 \def\hepa  {Helv.\wb Phys.\wB Acta}
 \def\ihes  {Publ.\wb Math.\wB I.H.E.S.}
 \def\ijmp  {Int.\wb J.\wb Mod.\wb Phys.\ A}
 \def\inma  {Invent.\wb math.}
 \def\jams  {J.\wb Amer.\wb Math.\wb Soc.}
 \def\joag  {J.\wB Al\-ge\-bra\-ic\wB Geom.}
 \def\joal  {J.\wB Al\-ge\-bra}
 \def\jodg  {J.\wb Diff.\wb Geom.}
 \def\jomp  {J.\wb Math.\wb Phys.}
 \def\lemp  {Lett.\wb Math.\wb Phys.}
 \def\maan  {Math.\wb Annal.}
 \def\maze  {Math.\wb Zeitschr.}
 \def\mpla  {Mod.\wb Phys.\wb Lett.\ A}
 \newcommand\npbF[5]  {\ {\sl #5}, \nupb\ {#1} [FS{#2}] ({#3}) {#4}}
 \def\nupb  {Nucl.\wb Phys.\ B}
 \def\phlb  {Phys.\wb Lett.\ B}
 \def\pnas  {Proc.\wb Natl.\wb Acad.\wb Sci.\wb USA}
 \def\ptps  {Progr.\wb Theor.\wb Phys.\wb Suppl.}
 \def\sebo  {S\'emi\-naire\wB Bour\-baki}
 \def\slnp  {Sprin\-ger\wB Lecture\wB Notes\wB in\wB Physics}
 \def\tams  {Trans.\wb Amer.\wb Math.\wb Soc.}
 \def\topo  {Topology}
 
 \def\A       {Algebra}
 \def\Ad     {{Amsterdam}}
 \def\alg     {algebra}
 \def\Be     {{Berlin}}
 \def\BIR    {{Birk\-h\"au\-ser}}
 \def\Ca     {{Cambridge}}
 \def\class   {classification }
 \def\compac  {compactification}
 \def\con     {conformal\ }
 \def\cua     {current algebra}
 \def\CUP    {{Cambridge University Press}}
 \def\dimn    {dimension}
 \def\furu    {fusion rule}
 \def\gefa  {Geom.\wb Funct.\wb Anal.}
 \def\GB     {{Gordon and Breach}}
 \def\ide     {identification}
 \newcommand{\iNBO}[7]{{\sl #7}, in:\ {\em #1} ({#5}), p.\ {#6}}
 \def\Infdim  {Infinite-dimensional}
 \def\inv     {invariance}
 \def\kzbe    {Knizh\-nik\hy Za\-mo\-lod\-chi\-kov\hy Ber\-nard equation}
 \def\KN      {Krichever\hy Novikov }
 \def\MD     {{Marcel Dekker}}
 \def\Modinv  {Modular invarian}
 \def\modinv  {modular invarian}
 \def\NH     {{North Holland Publishing Company}}
 \def\NY     {{New York}}
 \def\oa      {operator algebra}
 \def\parfu   {partition function}
 \def\PL     {{Plenum}}
 \def\Q       {Quantum\ }
 \def\qg      {quantum group}
 \def\qzn     {quantization}
 \def\Rep     {Representation}
 \def\SV     {{Sprin\-ger Verlag}}
 \def\syms    {sym\-me\-tries}
 \def\wzw     {WZW\ }
 \def\va      {Virasoro algebra}
 \def\WI     {{Wiley Interscience}}

 \end{document}